\documentclass[preprint]{elsarticle}
\usepackage{lineno,hyperref}
\usepackage{subcaption}
\usepackage{subfloat}
\usepackage{caption}
\usepackage{mwe}
\usepackage{changepage}
\usepackage{lscape} 
\usepackage[dvipsnames]{xcolor}
\PassOptionsToPackage{hyphens}{url} 
\usepackage{url}
\usepackage[hyphens]{url}
\usepackage{hyperref}  
\hypersetup{breaklinks=true}
\usepackage{tikz}
\usetikzlibrary{matrix,arrows,fit,shapes,decorations.markings,decorations.text,arrows.meta}
\usepackage{longtable}

\journal{Journal of Online Social Networks and Media}

\begin{document}

\begin{frontmatter}

\title{Fake news agenda in the era of COVID-19: Identifying trends through fact-checking content}
\tnotetext[mytitlenote]{This is a pre-print version of the full article available at \url{https://doi.org/10.1016/j.osnem.2020.100116}.}


\author[unifesp]{Wilson Ceron\corref{mycorrespondingauthor}}

\cortext[mycorrespondingauthor]{Corresponding author}
\ead{wilson.seron@unifesp.br}

\author[navarra]{Mathias-Felipe de-Lima-Santos}
\author[unifesp]{Marcos G. Quiles}

%endereços
\address[unifesp]{Federal University of Sao Paulo, Av. Cesare Mansueto Giulio Lattes, 1201, São José dos Campos, Brazil}

\address[navarra]{University of Navarra, Campus Universitario s/n, Pamplona, Spain}

\begin{abstract}

The rise of social media has ignited an unprecedented circulation of false information in our society. It is even more evident in times of crises, such as the COVID-19 pandemic. Fact-checking efforts have expanded greatly and have been touted as among the most promising solutions to fake news, especially in times like these. Several studies have reported the development of fact-checking organizations in Western societies, albeit little attention has been given to the Global South. Here, to fill this gap, we introduce a novel Markov-inspired computational method for identifying topics in tweets. In contrast to other topic modeling approaches, our method clusters topics and their current evolution in a predefined time window. Through these, we collected data from Twitter accounts of two Brazilian fact-checking outlets and presented the topics debunked by these initiatives in fortnights throughout the pandemic. By comparing these organizations, we could identify similarities and differences in what was shared by them. Our method resulted in an important technique to cluster topics in a wide range of scenarios, including an infodemic -- a period overabundance of the same information. In particular, the data clearly revealed a complex intertwining between politics and the health crisis during this period. We conclude by proposing a generic model which, in our opinion, is suitable for topic modeling and an agenda for future research.

\end{abstract}

\begin{keyword} 
Fact-checking\sep Fake news\sep COVID-19\sep Pandemic\sep Social Network\sep Twitter\sep Infodemic\sep Computational Method
\end{keyword}

\end{frontmatter}

%\linenumbers  %% <---- BUT HERE !

\section{Introduction}

On September 3rd, 2018, Fabiane Maria de Jesus, a young woman in her early thirtieth, could never imagine she would never return home after a church service. She died at the hands of a lynch mob driven by a series of vicious online rumors on the outskirts of Guarujá, a small coastal city in São Paulo, Brazil. Her death was motivated by the suspicion that Ms. Fabiane was connected to child abductions in the region thought to be linked to black magic rituals. However, investigations showed that the rumors were false. It all began with a post on Facebook that alerted local residents of a woman kidnapping children in the region to use them in witchery rituals. A tragic end for Ms. Fabiane shows the potential that disinformation has on social media\cite{Carpanez2018VejaGuaruja}.

Better known as fake news, the emergence of various forms of harmful content are many perils of social media that users can, even unconsciously, encounter in the online ecosystem\cite{Schapals2018FakeFacts}. The nature of information disorder of fake news makes it overlaps with other concepts that arose from the fabricated information that tends to mimic news content. In fact, fake news can take many shapes, it is therefore needed to distinguish between ‘misinformation’ and ‘disinformation.’ The first represents the claims that are false connected or inaccurate information, while the latter describes false information that is purposely spread to deceive people\cite{Lazer2018}. In the case of Ms. Fabiane, we can call that misinformation happened, once people shared the news that seemed accurate at the time but they later found it was made up. However, there are other situations where disinformation behaviors happen when an individual has intentional actions to purposely mislead people.
Thus, false information has real consequences, which makes it a relevant subject for study. This is especially true for troubled times that serve as nourishment for fake news\cite{Tandoc2018DefiningDefinitions}. The coronavirus pandemic has given a tremendous increase to this issue on social media networks. By taking root in a critical juncture filled with uncertainties and fears, the exploitation of the COVID-19 crisis has generated an excessive amount of information about a problem. Consequently, we also need to fight against the proliferation of false information about the topics related to the virus. This 'infodemic' ``can hamper an effective public health response and create confusion and distrust among people"\cite{WHO2020}. Furthermore, the decline in public trust in the media and the current political situation in a state of great flux around the world have even more notoriety to fake news\cite{Tandoc2019}.

There is no simple antidote to fake news. To face this problem, fact-checking became an important mechanism to fight falseness and information with the intent to harm\cite{Tandoc2019}. The rise of several fact-checking initiatives in recent years has shown the importance of debunking ``widely circulating claims online"\cite{Tambuscio2018NetworkFact-checking}.  However, conflicts in the logics that guide the operation of these sites have also been challenged by accusations of political bias\cite{Dobbs2012} and lack of clarity on what is being assessed by these initiatives, which shows the importance of understanding what topics actually are covered by these organizations. On the other hand, fact-checks can give a hint of what type of horrors lurk inside social networks.

Several studies have shown the development of fact-checking organizations in Western societies but little attention has been given to Global South\cite{Walter2019NewsApproach}. Special attention is necessary to address the dissemination of fake news in this region, where the rise of leaders like Jair Bolsonaro (Brazil), Daniel Ortega (Nicaragua), Rodrigo Duterte (Philippines), and Nicolas Maduro (Venezuela) reveals that autocratic populism can thrive in a wide range of environments to the propagation of rumors or false information, posing a global threat to democracy\cite{Flew2019,Goovaerts2020}. To address this gap, this study analyzes the trend based on a topic modeling approach to understand what was covered during the pandemic based on different fact-checking organizations. Therefore, in this study, we consider Brazil because even with the pandemic taking the media agenda in the world, the country has entered into a period of strong political instability, which makes the country even more susceptible to fake news campaigns\cite{Walter2019NewsApproach,Walter2020}. 

Drawing on a literature review of communication and computer science, we employed in this study a mixture of quantitative and qualitative methods including time-series analysis, topic modeling, and trend analysis. Through an original computational method inspired by Markov chains, we identify the ``fake news'' trends in Twitter accounts of Brazilian fact-checking organizations to further understand the rumors and false information that arose during the first half of the global health crisis. In this process, we also qualitatively analyzed the clusters to identify the topics that emerged in the time span defined by the authors.

Our contribution here is twofold to academic scholarship. First, we propose an innovative method that clusters topics based on an established time span. Second, we use this model to identify the topics that emerged in this extremely complex period known by an overabundance of the same information, also referred to as an infodemic, in a country that is facing a populist government, which makes it even more relevant. To show the usefulness of the proposed method, we focus our experimental investigation on tweets that were crawled of the timelines of the two major fact-checking initiatives in Brazil during the first half of the 2020 global health crisis.

Our approach tackles the abuse of social media along three dimensions: a better understanding of the debunks of the fact-checking initiatives during this period; the noisy points in the process by measuring the discrepancy between political and pandemic agendas; and also a correlation between the themes that were covered by these two initiatives to detect topic repetition and characterize the dissemination of false news information during this time frame.
 \\
\textbf{RQ1}\label{RQ1}: What are the topics that fact-checking initiatives covered during the pandemic? And how are these topics related between these two organizations?\\
\textbf{RQ2}\label{RQ2}:  How is the evolution of political topics intertwined with the pandemic crisis?\\
\textbf{RQ3}\label{RQ3}: What are the challenges to map topics in a period of overload of information on the same issue? 

The results obtained through the proposed method provide answers to the questions above and also our methodology is not restricted to this topic (fake news). The article is organized as follows. First, we propose a theoretical background to understand the importance of fact-checking initiatives and fake news in the media ecosystem. Second, we contextualize the situation of the pandemic in relation to the dissemination of false information and rumors to later introduce the relatedness of the Brazilian political agenda in this matter. Third, we describe our method to approach the problem. Fourth, we discuss the results relating to the evolution of false information and the influence of the political agenda during the health crisis. Last, we include some concluding remarks and further research issues.

\section{Related Work}

\subsection{Fact-checking: a task force to fight fake news}

The Internet has given everyone a voice, but many feel entitled to use these voices to spread false or misleading information. Consequently, the effect of propagation of misinformation, disinformation, and propaganda has recently been amplified due to the unprecedented increase in access to social media access. Ostensibly popularized as ‘fake news,’ and widely questioned in academia and the industry as it was used as a political weapon, this type of content predates the development of technology and social media and brings a range of meanings that have been associated with it \cite{Tandoc2018DefiningDefinitions}, such as destroying the credibility and trust of journalism as well as articulating and popularizing a political ideology. 

Previous studies have emphasized the strategies behind the dissemination of this content\cite{Tandoc2020WhenOutbreak}. The reasons behind go from money interest to political weapon, but the main purpose is to deceive the public \cite{Tandoc2019}. As has been previously reported in the literature, the efforts to misinform or disinform the public is understood by the financial motivation within the fake news chain from where it starts (upstream) to where and how it spreads (downstream). For example, research has provided evidence for the inadvertent funding granted to disinformation websites from programmatic advertising\cite{Bakir2018FakeSolutions,Braun2019FakeJournalism}. Furthermore, fake news relied on the platforms and their business models to gain traction by offering tools to actively promote dissemination and monetize content\cite{Lazer2018}.

Fake news has primarily drawn attention during the 2016 U.S. presidential election and became a constant problem, which made organizations look for ways to mitigate it\cite{Graves2018UnderstandingFact-Checking}. One of the most popular forms to address the prevalence of fake news has been fact-checking. The practice is defined by assessing the validity of claims made by institutions or public people ``to identify whether a claim is factual"\cite{Walter2020}. The idea behind this is to provide information to improve public discourse. Consequently, fact-checking organizations claim that they are nonpartisan and non-political entities. Thus, these initiatives establish a clear institutionalization to become more credible and acquire the professional prestige of the audience as a form to clarify their intention to become antidotes to misinformation.

Besides that, a close examination of scholarly literature has shown ambiguity about the efficacy of fact-checking \cite{Allcott2019,Walter2019NewsApproach,Walter2020}. While some studies have pointed out that political interest and ideology play a key role leading to a failure of fact-checking, others argue that debunks influence people's receptiveness to trust a source. Furthermore, the efficacy of fake news is contested by motivated reasoning and also preexisting beliefs that play a major role in determining the way information is processed. That happens because fact-checking organizations navigate in a ``tricky terrain"\cite{Walter2020}, discussing complex and often contradictory information whilst attempting to bring knowledge to the general public. Furthermore, there is an increasing polarization in online social media that has been gaining scholarly attention in recent years amid the changing political landscapes of many parts of the world. These polarized communities share a common narrative, which is empirically observed in the existence of echo chambers, stimulating the confirmation bias, i.e., the information confirming their beliefs even if containing false claims\cite{Prasetya2020AMedia}. Prior research suggests that in these groups, fact-checking corrections often fail to lessen misperceptions within an ideological group\cite{Nyhan2010WhenMisperceptions}. Also, studies documented numerous situations of a ``backfire effect'' in which these fact-checking debunks or corrections actually strengthen misperceptions amid these groups\cite{Nyhan2010WhenMisperceptions,Wood2018TheAdherence,Ecker2020TheFact-checks}.

In fact, political ideology plays a major role in determining how fact-checking is processed by individuals\cite{Ciampaglia2018FightingMisinformation}. Therefore, fact-checking reporting is usually accused of political bias, as readers believe that some candidates are likely to be more scrutinized than their counterparts. To overcome this, many organizations appeal to social-scientific approaches to selecting claims to avoid perceived bias\cite{Graves2018Aa}. Even so, important differences persist between topics that are covered by distinct initiatives, which reinforces the impression of bias.

However, many fact-checking initiatives exist with limited resources as they are yet small and employ few people that are insufficient compared to the scale of the fake news industry. Hence, there is a delay of approximately 13h between the consumption of fake news stories and its verifications. Thus, fact-checking lags behind misinformation and disinformation in terms of overall reach and response time\cite{Shao2016}. Still, to this date, the debate on how to improve the efficacy of fact-checking is far from resolved and all information is needed to come to a high-quality debunking decision.

Additionally, fact-checking has its peak during political events, such as elections or campaigns\cite{Coddington2014}. Scholarly studies have shown that people tend to share at a faster speed more false information on Twitter than other platforms, especially political topics\cite{Vosoughi2018a,Vargo2018}. In 2020, it was marked by a particular unique shock when the COVID-19 pandemic killed hundreds of thousands of people and forced all of us into social isolation, similarly to the period of elections when fact-checking is needed most\cite{Miller2015}. The health crisis has shown that individuals are more likely to be exposed to misinformation or disinformation\cite{Matos2020} and the importance of fact-checking to debunk these claims. Thus, the power of fake news encounters in its ability to penetrate social spheres \cite{Tandoc2018DefiningDefinitions}.

\subsection{Computational Social Science and Twitter}

Since the escalating notoriety of online social media platforms and the emergence of computational science, more and more researchers from a range of distinct areas have begun to examine social phenomena using large scale human-generated data computationally. Within this conjuncture, Twitter has become an indispensable social communication platform by offering many contrasting views and cultures on miscellaneous topics\cite{Boyd2012}. Indeed, a large number of existing studies in the broader literature took the opportunity of this platform to examine and get a better understanding of the interactions between individuals and their environment\cite{Boyd2012}.

Recent progress in various disciplines, such as mathematics and computer science, and particularly a subset of areas, such as network analysis, computational linguistics, and statistics have given a variety of tools to better understand social media content. By tracking, modeling, analyzing, and using mining techniques, scholars could grasp meanings out of the complexity of social media data\cite{Stieglitz2014SocialAnalytics}. The social network approach has given a valuable method to explain the way that information travels and how users put themselves at the intersections of routes where information flows\cite{Recuero2004TEORIAFotologs}. In fact, some studies argue that social networks are not always the most effective methods to identify collective action efforts\cite{Gonzalez-Bailon2016NetworkedMedia}. 

On the other hand, numerous methods have been developed to map the collective sentiment of tweets\cite{Pang2009OpinionAnalysis}, such as Naive Bayes, Support Vector Machine (SVM), linear classifier, and maximum entropy. These algorithms intend to classify tweets into positive, negative, or neutral\cite{Aston2014TwitterPerceptron}. Yet, another study approached the sentiment of tweets based on hashtags and emojis by manually labeling 12 categories that were then used to identify the users' perceptions of these messages\cite{Davidov2010EnhancedSmileys}.

Another prominent area of study in computational social science is the trend and time-series analysis. A recent study analyzed in Facebook the evolution of rumors that related the pandemic outbreak to the rollout of 5G broadband technology using the time-series analysis to understand the evolution of this disinformation throughout the time analyzed\cite{Bruns2020CoronaFacebook}. Despite this study, false information has spread to all social media platforms, and Twitter has become an increasingly popular social network to study the evolution of the pandemic. To illustrate this, deniers of the pandemic used the hashtag \#FilmYourHospital on Twitter to encourage users to visit hospitals and hinder the work of physicians and nurses in the middle of a health crisis to take pictures to ``prove'' that the COVID-19 pandemic is an elaborate hoax\cite{Gruzd2020GoingTwitter}. In fact, several methods to predict the trends of news topics on Twitter have emerged in recent years\cite{Karami2020}. For example, prior work has applied trend analysis methods, such as the Moving Average Convergence-Divergence (MACD) indicator, to understand the news topics that were trending at the moment and could potentially become popular on Twitter\cite{Lu2012NumericalBodies}.  

Similarly, topic modeling became increasingly popular on social networks. Many sources give methodological guidance on how to apply LDA topic modeling on news articles, for instance, on temporal analysis of climate change articles published between 1997 and 2016\cite{Keller2020NewsTopics}. However, the application of the LDA algorithm in tweets was reported in the literature that may not work well because the text corpus is short\cite{Zhao2011ComparingModels,Hong2010EmpiricalTwitter}. To overcome this, a study has used a fetched category from news articles to compare the classification of tweets using Naive Bayes and SVM algorithms, and concluded the latter is more effective\cite{Dixit2018}. Other studies proposed to aggregate tweets as a single document. An example of this is the technique to group together tweets occurring in the same user-to-user conversation and use a combination of Latent Dirichlet Allocation (LDA) and the Author-Topic Model (ATM) to determine topics in tweets\cite{Alvarez-Melis2016TopicConversations}. 

In the same vein, existing studies in the broader literature have examined the applicability of LDA in Twitter messages to detect fake news. By analyzing the main topics of the messages with reliable sources, such as magazines and news portals, it helped to identify false information disseminated in tweets\cite{deSouza2020AMedia}. Another method introduced was the combination of sentiment analysis with topic modeling. In it, the authors extracted the sentiment from replies or retweets to analyze the general opinion on topics and assist LDA in the classification of fake news\cite{Chatterjee2016TwitterCredibility,Vohra2018DetectionMedia}.

A recent study used sentiment analysis and LDA to analyze tweets and concluded that there is a wide range of automated accounts that were disseminating content on Twitter. For instance, in a set of 29 million tweets, the authors found that $21\%$ of the accounts were bots and, consequently, $30\%$ of these tweets. However, it is not clear whether the implications of bots are always negative on social networking sites, once there are automated accounts that share reliable information\cite{Liu2019}.

Other methods were also proposed to detect false information on social media content without topic modeling algorithms, such as Principal Component Analysis  --  PCA\cite{YanZhang2016ADifferences} and Statistical Relational Learning -- SRL\cite{Wu2018TowardMisinformation}. PCA is a technique to extract features and in general, used to eliminate irrelevant information. In this case, it applied to identify rumors and false content. In the same vein, SRL combines machine learning and data mining to detect noisy, missing, or partially observed information\cite{deSouza2020AMedia}.

Despite these studies showing good results, a number of questions regarding the evolution of fake news agenda in time remain to be addressed. Furthermore, there is no solution to curb fake news at the moment. As technology spawned the dilemma of fake news, it is tempting to assume that technology can solve it. In this context, scholars proposed a range of strategies to curb the dissemination of false information on social media, for instance, focusing on the role of network polarization\cite{Tambuscio2019Fact-checkingSociety}. In addition, artificial intelligence (AI) solutions have been particularly effective in detecting and removing dubious or undesirable content online. On the other hand, scholars raised concerns about the accuracy and transparency of AI by placing responsibility for technology to detect and tag what is false content on social media platforms\cite{Skolkay2019ASolutions}. Methodology-wise, combining computational methods and human analysis provides a more effective way to explore such information in social media\cite{Meyer2004ImpfgegnerImpfskeptiker}.
%%%%%%%%%%%%%%%%%%%%%%%%%%%%%%%%%%%%%%%%%%%%%%%%%%%%%%%%%%%%%%
%%%%%%%%%%%%%%%%%%%%%%%%%%%%%%%%%%%%%%%%%%%%%%%%%%%%%%%%%%%%%%%%%%

\subsection{From a health crisis to a political firestorm: How Brazil is handling the outbreak}

The newness of the COVID-19 virus means that no one has yet been able to understand the etiology of the pandemic. This has caused many uncertainties and fears in the population. To make matters worse, as the COVID-19 pandemic made landfall in every continent, fake news ramped up on just about every platform, causing widespread fear and confusion. The furor over fake news has intensified long-standing concerns in this fragmented media environment that is seen in society today, with many sources of information. Furthermore,  online rumors and political lying are other concerns that have exacerbated this fake news scenario \cite{Graves2018UnderstandingFact-Checking}.

This mix of facts, fears, rumors, and speculations has characterized the pandemic and provided fertile ground for the dissemination of misleading information and cybercrime on the Internet\cite{Nguyen2020,Vraga2020}. Considering that fake news that ``spreads faster and more easily than''\cite{WHO2020} a virus, this became a concern during the pandemic. For all that, the World Health Organization (WHO) has declared that false information shared on social media about COVID-19 is an ``infodemic" that must tackle alongside the pandemic itself and the concern over the problem is global.

But apart from this issue, some heads of state stand apart for their continued denials of the threat the pandemic poses. To illustrate this, leaders of Belarus (Alexander Lukashenko), the USA (Donald Trump), and Brazil (Jair Bolsonaro) use their position to deny the global movement to fight the pandemic and described the illness as a ``little flu,'' or promoting a medicine (hydroxychloroquine) as a cure for the illness, despite growing scientific evidence against the efficacy of the malaria drug\cite{Schipani2020, Friedman2020}. In a scenario where official information is understood as unreliable, the climate is set for the viral dissemination of unproven rumors and speculations \cite{Larson2020}. 

In this environment, not everything is false, some ``emerging, albeit unverified, information might be valuable, and deleting it would cause harm"\cite{Larson2020}. In fact, the effects of this factually dubious content produced with no regard for accuracy or fairness that mimics the format of journalism in the public remain to be addressed \cite{Lazer2018}. On the other hand, it also made more evident than ever the fundamental importance of fact-checking initiatives in periods of crisis, mainly where politicians made a complicated landscape for citizens. In Brazil, the political scenario has created many uncertainties and fears in the population. The systematic violation of access to information and transparency during the coronavirus pandemic made by the Brazilian Federal Government has failed to mitigate the impacts of the pandemic on the population \cite{Matos2020,ARTICLE192020}. Consequently, major legacy media organizations teamed up to provide transparency of daily coronavirus death tolls and infection rates to citizens\cite{FolhadeS.Paulo2020}.

In the same vein, fact-checking initiatives play a crucial role in debunking and verifying information that is disseminated in social media in Brazil. Some organizations in the country target politicians exclusively but others focus mainly on correcting mistakes in the media or even in niche areas, such as fact-checking of religious content posted by Christian personalities, especially politicians and influencers in social media platforms \cite{Batista2020}. Among the most recognizable fact-checking organizations in Brazil are \textit{Agência Lupa} and \textit{Aos Fatos}\cite{Dourado2020AgenciaINTERNET,Moreno2019Factck.br:News}.  Besides being the major producers of fact-checking stories, these organizations are committed to the transparency standard of the International Fact-Checking Network (IFCN), an alliance that abides by promoting excellence in fact-checking\cite{IFCNInternationalPoynter}. Whereas most fact-checkers in the United States have ties to a news outlet, internationally fewer than half do\cite{Stencel2016GlobalLab}, what it is exactly the case of these two organizations that have their entire operation focused on debunking information propagated in social media platforms. Furthermore, these two initiatives have a huge impact on social media, for example on Twitter, \textit{Agência Lupa} has more than 181K people following their content, while \textit{Aos Fatos} counts with 238K followers (as of November 2020). 

Our analysis deals only with these two organizations that came about due to the huge demand for rigorously verified information, with public and transparent data that gained evidence in the last years. Thus, this study wants to uncover features and issues that were debunked during the pandemic to test an original method to evaluate the trends of fake news topics on Twitter based on a Markov-inspired model. A comprehensive description of methods and data has been provided to support our results in the next section.

\section{Data and methods}

Beyond electoral impacts, it is still restricted what we know about the effects of fake news content in periods of crisis, similar to the COVID-19 health crisis. From increasing cynicism and apathy about the health crises to encouraging extremism with false statements, this article studies the topics that have risen up the agenda during this period. There exists little evaluation of the trends of fake news in these regards, and also a computational method to identify these topics properly. To address this gap, this study proposes a comprehensive method that provides a nicely integrated solution to identify these trends in a time-series analysis.

\subsection{Data description}

To investigate the evolution of fact-checking trends during the pandemic and understand how the political agenda has influenced the fake-news agenda, we analyzed tweets from Brazilian fact-checking accounts. Instead of a simple topic modeling, we propose a method that clusters the topics and its evolution in an established time. To address this phenomenon, this article relies on the open Twitter application programming interface (API) connection to crawl the timelines of the two major fact-checking agencies in Brazil, \textit{@aosfatos} and \textit{@agencialupa}. In total, these accounts account for 5115 tweets from January to July 2020.

We concentrate on two fact-checking organizations, and we do not investigate the numerous other smaller initiatives existing in the platform because the number of tweets produced by them during the period was not enough to be considered in this study. Nevertheless, because our samples include activities of the most prominent fact-checking organizations, we can draw consistent inferences about our method and the questions proposed by this research piece.

We retrieved the tweets and metadata related to these accounts to assemble the pool of data for our computational method. This choice is made because these organizations posted the most tweets related to fact-checking content over the course of the period appraised. Also, as mentioned before, these organizations are well embedded in a set of historically rooted values, traditions, and norms that shaped the fact-checking process. Other initiatives were considered but their number of tweets was not enough to compute effective temporal features. After devised compilation and cleansing, data is further used in our computational model.

\subsection{Data preparation}

Literature has shown different approaches for topic modeling, such as Latent Dirichlet Allocation – LDA\cite{Blei2003LatentJordan} and Latent Semantic Analysis – LSA\cite{Scott1990IndexingAnalysis} and also best practices that yield more coherent topic-bags\cite{Martin2015MoreApproach}. However, these models are independent of time, and to overcome this issue, other algorithms are used to obtain this temporal information, such as DDTM (Discrete-time Dynamic Topic Model), CDTM (Continuous-time Dynamic Topic Model), ToT (Topics over Time), and TAM (Trend Analysis Model). However, these models also deal with the limitation of the complexity of variational inference or there is a need to delimit the number of topics beforehand\cite{Lu2012TrendTwitterb}. To address this gap, we derive a novel algorithm that analyzes tweets in the fortnightly tide by clustering the output inspired by Markov chains\cite{Coviello2014ClusteringHEM}. We made the decision to divide into fortnights because that encompasses an ideal number of functional units that can be formed as a community (or cluster).

Due to the nature of Twitter data, there is a lot of noise among words. Preprocessing of tweets is still a popular technique to help in this procedure. It is a common task for text normalization and almost every task related to natural language processing (NLP) to apply some kind of tokenization, stemming, and lemmatization. In our case, the tokenization is performed in the context of graphs that were generated to represent the tweets. The purpose of stemming and lemmatization is to lessen inflectional forms and sometimes derivationally similar modes of a word to a common base form. However, the Portuguese language has its own particular characteristics. Lemmatization considers the syntactic classification of words, presenting, for instance, different lemmas for a verb or a noun (in the same word family).  In English, this nuance is different, once the same lemma can assume multiple syntactic categories. Instead, in Portuguese, and other romance languages, it is of predominant importance, and changes the meaning of the sentence. In practice, many exceptions to these same rules frustrate the goal of achieving an accuracy close to 100\% in most of the romance languages\cite{Rodrigues2014Lemport:Portuguese}.

An example, the word campeão (champion) in singular and campeões (champions) in the plural to extract the end of these two words includes many exceptions that do not necessarily fit the typical case for one option or another and result in an impossible task for computers to perform. Stemming is easier to implement, usually refers to a crude heuristic process that chops off the ends of words. However, it discards possibly valuable information, by making it practically difficult to identify, for instance, a noun from a verb in its stemmed form\cite{Rodrigues2014Lemport:Portuguese}.  Thus, due to the complexity of this task for the Portuguese language and at the risk of losing sight of the main point of the tweet, our algorithm does not include lemmatization and stemming processes in the pre-processing phase.

On the other hand, we decided to first map all threads and treated each of them as a single tweet to encompass the same topic. To be able to achieve it, the tweets that were posted in the time interval of fewer than two minutes were aggregated into a single one. Known as Twitter threads, it is a series of connected tweets from one account and usually contains the same sort of elements. This was an important step to enrich the vocabulary of a certain topic and guarantee that neither the content would remain separated. Secondly, a pre-processing step consisted of eliminating stop words and punctuation signs. We additionally removed the ``RT" from the retweets, mentions, and hashtags. We also evaluate the impact of different models of lexicons for the Portuguese language and the effect of pre-processing techniques for the task and mapped the common words that are not relevant and appear in our dataset to be removed.

LDA and LSA algorithms were used to help in the process, as these algorithms use probabilistic models to extract latent bags of words or topics in a document. We run a number of times these algorithms, which did not present good results in the clusterization process, but they served along with our model in the optimization of the pre-processing phase. Likewise, after evaluating and deleting the terms and expressions that are common by these fact-checking agencies, such as ``We verified" and ``This is false," we prepare our data for the application of our method based on graphs. Other specific and popular terms in this specific context, such as coronavirus and COVID-19, were also removed to help in the clustering process. To illustrate the preprocessing model, a tweet in our dataset was like this one:\begin{itemize}
   \item \textit{NO AR. OMS não orientou evitar sexo com animais para se prevenir do coronavírus. (LIVE. WHO did not guide to avoid sex with animals to prevent from coronavirus, in English) \href{https://www.aosfatos.org/noticias/oms-nao-orientou-evitar-sexo-com-animais-para-se-prevenir-do-coronavirus/}{https://aosfatos.org/noticias/oms-n...}  $\# CoronaVirusFacts$}
\end{itemize}
After the preprocessing, it should have been coded in that way:
\begin{itemize}
   \item \textit{OMS ORIENTOU EVITAR SEXO ANIMAIS PREVENIR (WHO GUIDE AVOID SEX ANIMALS PREVENT, in English) 
}
\end{itemize}

\subsection{Markov chains and topic modeling}

Markov chains are widely applied in various fields, such as economics, finance, sports, and even are applicable to on our daily routine, such as typing in smartphones, where the device predicts what one is going to type next. However, the model has not ``yet been widely implemented in communication research''\cite{Vermeer2020TowardApproach}. Unlike the conventional view of classifying tweets, like LDA, Markov chains use a stochastic approach that is formally equivalent to the equations for steady-state probabilities.

	\begin{figure}[h!]
	\centering
		\begin{tikzpicture}
		\tiny
		\tikzset{ scale=0.6,auto=right}
		\tikzset{node/.style={circle,draw=black,fill=white!20, minimum size=2cm}}
		\tikzset{edge/.style = {->,> = latex', line width=1pt,draw=black}}
     \tikzset{el/.style = {inner sep=2pt, align=left, sloped} }

			\node[node] (n1) at (-14,9) {$\textbf{OMS}$};
			\node[node] (n2) at (-10,14) {$\textbf{ORIENTOU}$};
			\node[node] (n3) at (-6,9) {$\textbf{EVITAR}$};
			\node[node] (n4) at (-2,14) {$\textbf{SEXO}$};
			\node[node] (n5) at (2,9) {$\textbf{ANIMAIS}$};
			\node[node] (n6) at (6,14) {$\textbf{PREVENIR}$};		

      \draw[edge] (n1) edge node[el,above]{$w_1=1$} (n2);
      \draw[edge] (n2) edge node[el,above]{$w_2=1$} (n3);
      \draw[edge] (n3) edge node[el,above]{$w_3=1$} (n4);
      \draw[edge] (n4) edge node[el,above]{$w_4=1$} (n5);
      \draw[edge] (n5) edge node[el,above]{$w_5=1$} (n6);
			%\draw[edge] (n5) -- (n1);
		 
		\end{tikzpicture}
		\caption{Example of a directed graph based on the pre-processed tweet: ``\textit{OMS ORIENTOU EVITAR SEXO ANIMAIS PREVENIR}''} \label{fig:graph}
	\end{figure}
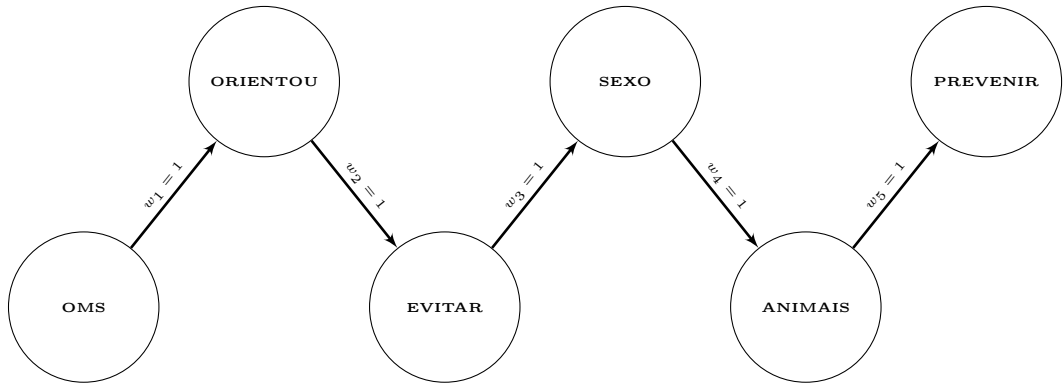

Here we develop a network that is inspired in a Markov chain\cite{Tierney1991ExploringChains}, our method that generates a graph $G = (V, E)$ that is composed of vertices $V = {v_1,v_2,v_3…v_n}$ that represent the words of a tweet and the edges $E = {e_1,e_2,e_3…e_m}$ are the connections between these words. As each tweet, with $n = |V|$ vertices and $m = |E|$ edges, involves a sequence of words, our graph here is directed, thus, it respects the text structure. Furthermore, each E has a weight $W = {w_1,w_2,w_3…w_m}$ that represents how many times these two words were connected in the dataset, as shown in figure \ref{fig:graph}.The importance of directed graphs in our methods lies in generating the right knowledge to form the sequence of processed words in the tweets. For example, ``Flavio'' (subject) and ``steal'' is not the same as ``steal'' and ``Flavio'' that is the reason a directed graph is necessary to maintain the ideas that a sentence shares. Similarly, it is the relation in Twitter, one person follows someone that does not necessarily follow the same person back\cite{Boyd2012}.

Besides that, cycles were not considered in this method, once words that are feeding themselves do not bring extra information to our topic modeling approach. Weighted graphs are a more informative graph, which tells the probability of visiting the next node or maybe the distance of a node to another or maybe the dependency of one node over the other\cite{Tierney1991ExploringChains,Geyer1992PracticalCarlo}. Once the graph is built, a community detection method highlights the main communities of the network and, consequently, the topics.
The graph was constructed using tweets within a certain time window. Specifically, after a careful investigation of distinct time resolution (days, weeks, and a month), we observe that, for the scenario considered in this investigation, a time window of two weeks, i.e., 15 days. This number was decided based on experiments to obtain a representative sample of tweets of each organization that could be evaluated by our algorithm. In fact, it was also noted through our experiments that the major topics of the dataset remained in this time span. The method is then applied for tweets within each interval, thus, each graph was composed of the words $(V)$ and the connection between them $(E)$. To provide an intuitive understanding of the algorithm proposed here, we described the steps conducted below: 
\begin{itemize}
    \item After pre-processing, each tweet that is loaded, $v_i$ are added to the graph. In case, a word is already on the graph, it is not added. 
    \item  The tweet is read from left to right (the same way that we read) and for each pair of words that are subsequent in the dataset, an $e_i$ is added with weight 1. Again, in case the edge already exists, no other edge is added, but the weight increases by 1. 
    \item This process happens until all tweets are loaded in the graph. 
\end{itemize}
As a result, the obtained graph is a subset of the dataset that conforms to a time span. Thus, simple, cost-effective measures could be found, such as the degree of any vertices and the weights of any edges, which reveals some important information about the dataset. To put it another way, in many fortnights, the top five heaviest edges were related to the main topics found during the study period. After that, the graph is clustered, which we used an edge-clustering framework that can group the heaviest edges into bundles to reduce the overall edge crossings.

Furthermore, we ran our method a couple of times to detect possible words that could be merged in the pre-processing phase. For instance, in our initial analysis, we identified that the edges with higher weight represent compound names, such as city or states’ names (Rio de Janeiro) or important figures (Jair Bolsonaro or Sérgio Moro). Indeed, this was an important step to return to our data preparation. In many cases, pre-processing is not a single operation. Instead, pre-processing is an entire pipeline composed of many phases to achieve an optimal formulation.

In our graph, we relied on a Python library \textit{igraph} that provides different types of clustering methods. In an echo of scholarly literature, we selected the walktrap algorithm, as it delivers the correct number of communities regardless of network sizes\cite{Yang2016ANetworks}. Walktrap approach based on random walks, i.e., run unsystematically the graph, thus, the walks are more likely to stay within the same community because there are only a few edges that lead outside a given community, which result in more precision in the topic modeling. Consequently, our graph could identify clusters of tweets that were closely linked with each other, thus determining a topic. For each graph, we qualitatively analyzed the clusters to observe the different topics. Regarding community detection, any method that is suitable for directed networks would work in our case. As a methodological choice, walktrap is a method that we are used to and we have a certain domain about it. Other community detection methods would work as well in our algorithm. 

Our time-series analysis divided the period into fortnights and generated the clusters of the five major clusters for each individual period to obtain meaningful information on the evolution of the topics covered by the organizations during the time assessed. Since classifying the clusters would involve a considerable manual effort, we prioritized the annotation of the five major clusters, once those groups usually represent a larger number of tweets. We also limited the number of topics to the five major communities, as we would like to visualize it through figure \ref{fig:chart} and \ref{fig:topics}, which were produced using the \textit{ggplot2} library in R. Furthermore, topics that were closely related or too small were not considered, thus reducing the number of topics in a certain time span. 

\begin{figure}
  \begin{subfigure}[b]{0.5\textwidth}
  \captionsetup{justification=centering}
    \includegraphics[width=\textwidth]{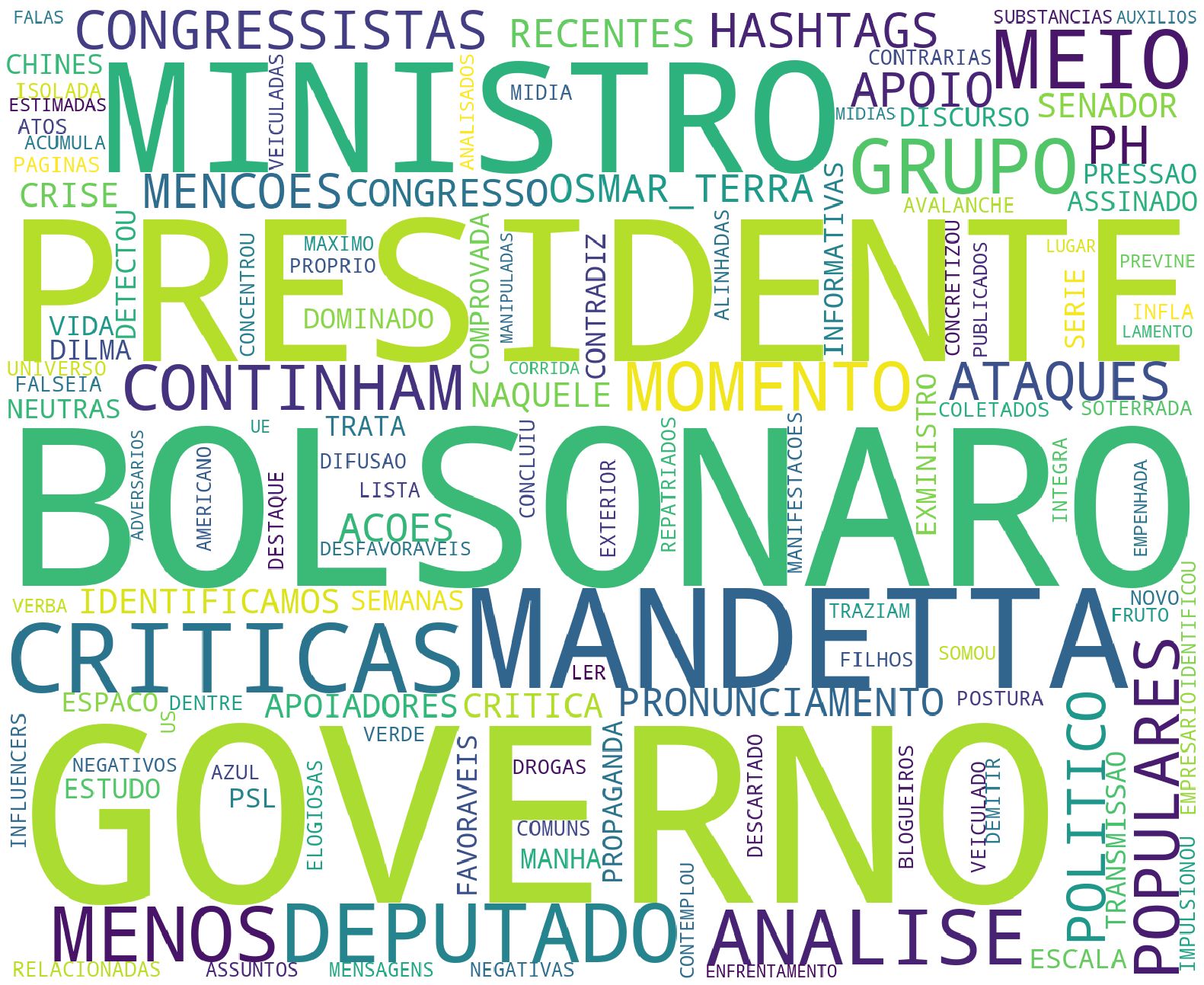}
    \caption{Topic: Federal Government - COVID19\\\textit{Aos Fatos}-- Fortnight 7}
    \label{fig:}
  \end{subfigure}
  \begin{subfigure}[b]{0.5\textwidth}
    \includegraphics[width=\textwidth]{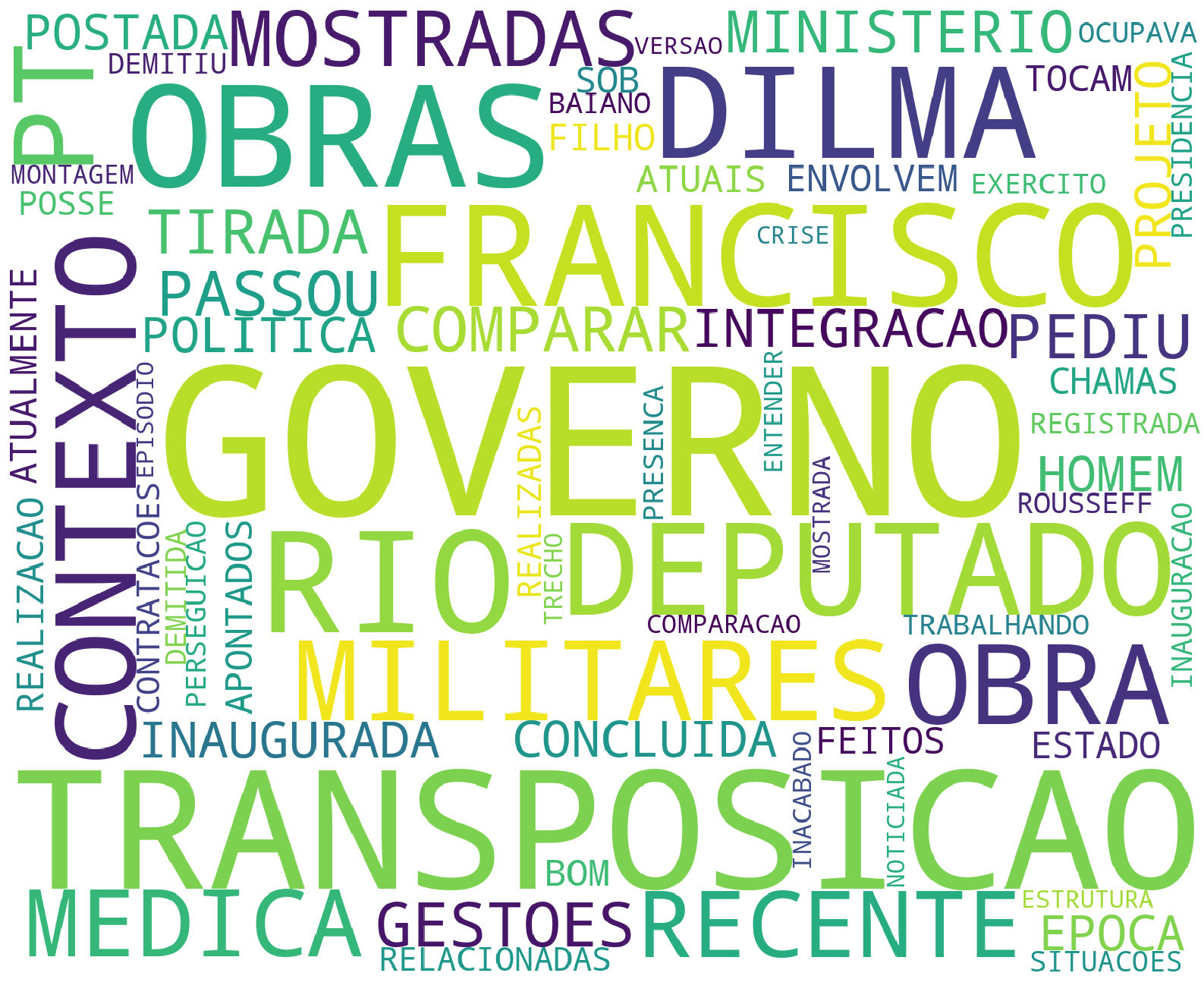}
      \captionsetup{justification=centering}
    \caption{Topic: Federal Government\\\textit{Aos Fatos} -- Fortnight 12}
    \label{fig:}
  \end{subfigure}

  \begin{subfigure}[b]{0.5\textwidth}
    \includegraphics[width=\textwidth]{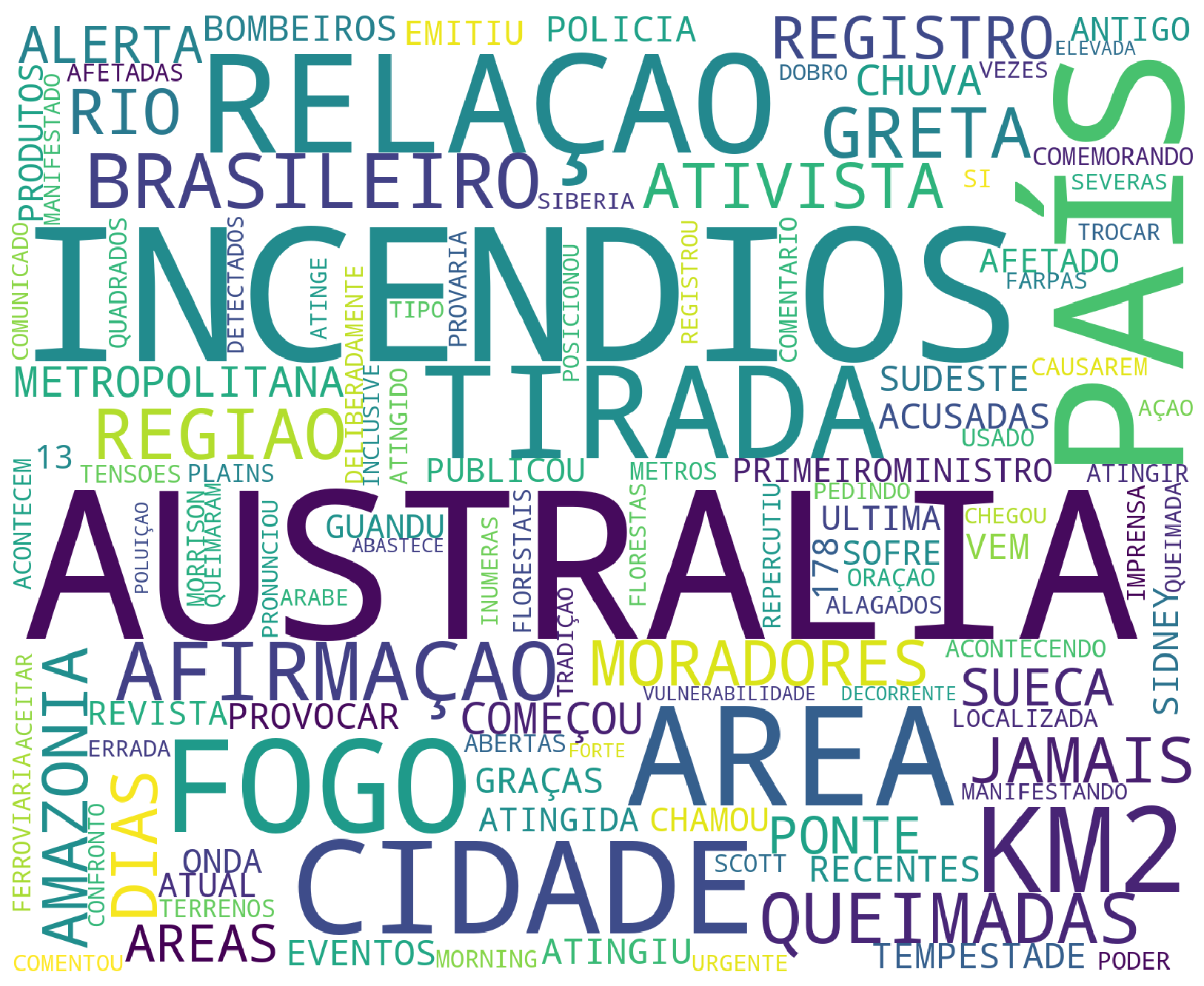}
  \captionsetup{justification=centering}
    \caption{Topic: Environment\\\textit{Agência Lupa} -- Fortnight 1}
    \label{fig:}
  \end{subfigure}
  \begin{subfigure}[b]{0.5\textwidth}
    \includegraphics[width=\textwidth]{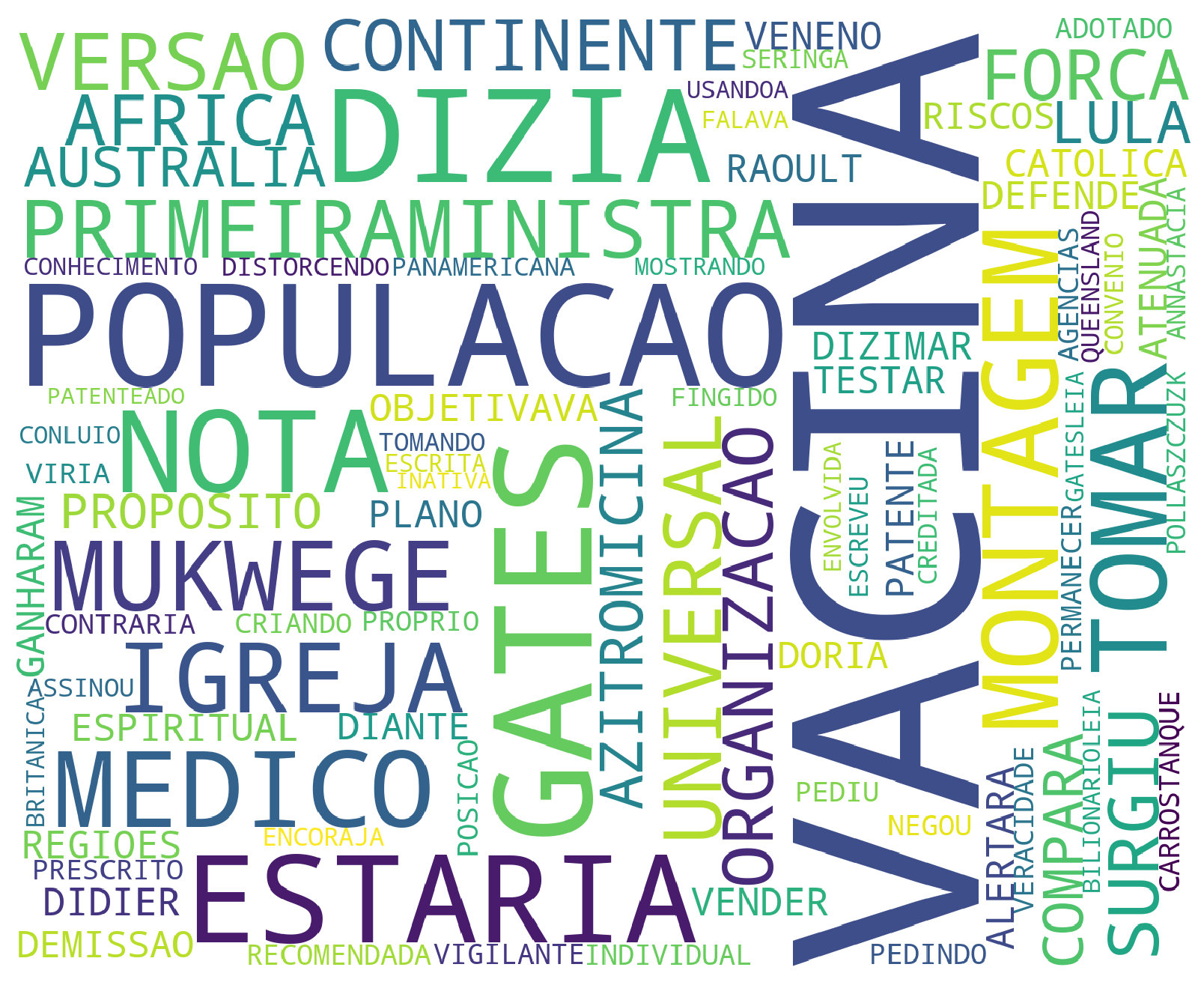}
  \captionsetup{justification=centering}
    \caption{Topic: Prevention - COVID19\\\textit{Agência Lupa} -- Fortnight 12}
    \label{fig:}
  \end{subfigure}
\caption{Random word clouds to show the communities generated by the computational method}\label{fig:wordcloud}
\end{figure}

Another important point is that to better visualize the communities, we decided to adopt a word cloud to represent each community. In it, the words represent the vertices and their weight. Hence, it gives greater prominence to vertices with higher weight, the ones that were more frequently used in that cluster. Lastly, it is worth mentioning that the smallest clusters were threads or one single tweet that was not relevant for this study. Thus, we created word clouds of the most connected vertices (filtered above 10 degrees), as shown in figure \ref{fig:wordcloud}.

We also observe in our results the presence of clusters in specific languages. Although the fact-checking organizations use Portuguese in their tweets, some retweets were in English or Spanish. As the process of creation of the graph connects related words, it is natural that words in the same language present more connection between them than in other languages. Thus, it is easily detected as a cluster by our algorithm.
One disadvantage of the most traditional methods, such as LSA and LDA that are probabilistic methods (related to the frequency of a word in relation to its neighbor), is that the precision depends on the size of the dataset and lexical diversity. The method here proposed is instead designed to be not influenced by these factors, once it considers the connection of the words contained in tweets in relation to the dataset. The quantity of topics is also another strength as it does not need to be previously established. However, by adopting a rigid clustering system, there is a risk that the clusters formed will be a thread or a single tweet. This level of adjustment also allows better control of the topics found. 
In the process of labeling our clusters, we identified 23 codes in the whole dataset, which includes Federal Government (topics related to the Presidential Government), Politics (in general), Health, and also topics that were related to COVID-19 associated with other areas like COVID-19 – Dissemination or COVID-19 – Treatment (see more in Table \ref{tab:agencialupa} and \ref{tab:aosfatos}). By applying the method here described, we could identify the topics and the patterns that were covered by fact-checking agencies during the first half of 2020 and were described next.

\begin{table}[]
\hspace*{-2.0cm}
\tiny
\begin{tabular}{|l|l|l|c|c|c|}
\hline
Timeframe    & Topics                                                        & Code                         & Cluster (\%) & Vertex & Edge \\ \hline
Fortnight 1  & Iran - USA - General                                          & International Affairs        & 6.95         & 148    & 246  \\ \hline
Fortnight 1  & Australia - Bushfire - Fire                                   & Environment                  & 3.05         & 65     & 100  \\ \hline
Fortnight 1  & GeniuX - Product - Medicine                                   & Health                       & 2.39         & 51     & 58   \\ \hline
Fortnight 1  & Water Rio - CEDAE                                             & Public Policy                & 2.21         & 47     & 63   \\ \hline
Fortnight 2  & SP - Crackland - Investment - Government                      & Politics                     & 11.66        & 203    & 334  \\ \hline
Fortnight 2  & Brumadinho - Disaster                                         & Disaster                     & 13.33        & 58     & 65   \\ \hline
Fortnight 2  & Water Rio - CEDAE                                             & Public Policy                & 2.41         & 42     & 48   \\ \hline
Fortnight 2  & Non-existent employee - Paiva - President                     & Federal Government           & 2.01         & 35     & 38   \\ \hline
Fortnight 2  & Animal - Heineken - Dog-baiting - Sponsor                     & Business                     & 1.95         & 34     & 37   \\ \hline
Fortnight 3  & Economy - Bolsonaro - Minister - Guedes                       & Federal Government           & 5.83         & 136    & 198  \\ \hline
Fortnight 3  & Pandemic - China - Virus - Flight                             & Dissemination - COVID19      & 3.77         & 88     & 117  \\ \hline
Fortnight 3  & Sickness - Spread - Market                                    & Transparency - COVID19       & 2.14         & 50     & 52   \\ \hline
Fortnight 3  & Economy - ANEEL - Light                                       & Economy - COVID19            & 1.93         & 45     & 50   \\ \hline
Fortnight 3  & Fennel - Tea - Throat                                         & Prevention - COVID19         & 1.93         & 45     & 53   \\ \hline
Fortnight 4  & IPVA - IPTU - Fuel - ICMS                                     & Economy                      & 7.83         & 113    & 149  \\ \hline
Fortnight 4  & Councillor - Journalist - Population                          & Media                        & 7.07         & 102    & 126  \\ \hline
Fortnight 4  & Carnival - Glitter - Environment                              & Leisure                      & 6.51         & 94     & 131  \\ \hline
Fortnight 4  & Pope Francis - Manifestation - Lula                           & Politics                     & 3.26         & 47     & 56   \\ \hline
Fortnight 4  & Alcohol Gel - Wash hands                                      & Prevention - COVID19         & 2.01         & 29     & 29   \\ \hline
Fortnight 5  & Vaccine - Alcohol Gel - Wash Hands - Breathalyzer             & Prevention - COVID19         & 5.28         & 160    & 276  \\ \hline
Fortnight 5  & Carnival - Agglomeration - Isolation                          & Dissemination - COVID19      & 4.71         & 143    & 172  \\ \hline
Fortnight 5  & Cancer - Skin Cancer - Sun                                    & Health                       & 2.54         & 77     & 92   \\ \hline
Fortnight 5  & Bolsonaro - Deputies - Manifestation                          & Federal Government           & 1.62         & 49     & 74   \\ \hline
Fortnight 6  & Information - Vaccine - Sickness - Pandemic                   & Transparency - COVID19       & 15           & 512    & 991  \\ \hline
Fortnight 6  & Governors - Secretary - President - States                    & Politics - COVID19           & 4.39         & 150    & 213  \\ \hline
Fortnight 6  & Closure - Israel - Cuba - Sweden - President                  & Dissemination - COVID19      & 2.75         & 94     & 102  \\ \hline
Fortnight 6  & Emergency Aid - Numbers - Deputies                            & Economy - COVID19            & 1.82         & 62     & 67   \\ \hline
Fortnight 6  & Gas - Gas Station - Harmful Products                          & Prevention - COVID19         & 1.82         & 62     & 68   \\ \hline
Fortnight 7  & Flight - COVID 19 - Cases - Numbers                           & Transparency - COVID19       & 4.68         & 173    & 276  \\ \hline
Fortnight 7  & Governor SP - Hospitalized - Police                           & Local - COVID19              & 3.84         & 142    & 224  \\ \hline
Fortnight 7  & Hospitals - ICU - Hospital Beds - Health Secretariat          & Hospital - COVID19           & 2.19         & 81     & 120  \\ \hline
Fortnight 7  & Hydroxychloroquine - Treatments - Eficacy                     & Treatment - COVID19          & 2.03         & 75     & 106  \\ \hline
Fortnight 7  & Deputies - President - Chambers - Former President            & Politics - COVID19           & 1.95         & 72     & 109  \\ \hline
Fortnight 8  & Pandemic - COVID19 - Masks - Dissemination                    & Dissemination - COVID19      & 7.89         & 347    & 606  \\ \hline
Fortnight 8  & Justice - Sergio Moro - Minister - Bolsonaro                  & Federal Government           & 3.75         & 165    & 235  \\ \hline
Fortnight 8  & Number - Cases -  Notaries - Data                             & Transparency - COVID19       & 3.55         & 156    & 241  \\ \hline
Fortnight 8  & Amazonas - Empty Coffins - Municipalities                     & Local - COVID19              & 2.98         & 131    & 169  \\ \hline
Fortnight 8  & Open Coffins - Empty Coffins - Alleged - Recording            & Politics - COVID19           & 1.68         & 74     & 85   \\ \hline
Fortnight 9  & Respiratory Illness - Deaths - Numbers                        & Transparency - COVID19       & 6.28         & 262    & 502  \\ \hline
Fortnight 9  & Against - Globo Group - CNN Brazil - Journalists - Interviews & Media - COVID19              & 2.92         & 122    & 164  \\ \hline
Fortnight 9  & Hospital - Isolation - Recording - 99\% cured                 & Hospital - COVID19           & 2.73         & 114    & 135  \\ \hline
Fortnight 9  & Data - Civil Registry - Transparency - Portal                 & Transparency - COVID19       & 2.37         & 99     & 153  \\ \hline
Fortnight 9  & Cellphone - Bolsonaro - PGR - STF - Seize                     & Federal Government           & 1.84         & 77     & 98   \\ \hline
Fortnight 10 & Isolation - Positive Test - South Corea - COVID19             & Dissemination - COVID19      & 13.85        & 474    & 955  \\ \hline
Fortnight 10 & Vaccine - Test - Studies - OMS - Ineficacy - Quarentine       & Prevention - COVID19         & 1.64         & 56     & 68   \\ \hline
Fortnight 10 & Pazuelo - Interim Minister - Saude                            & Federal Government - COVID19 & 1.37         & 47     & 53   \\ \hline
Fortnight 10 & Car - Parking lot - Macae - PT - Worker's Party               & Politics                     & 1.34         & 46     & 50   \\ \hline
Fortnight 11 & Pandemic - Symptoms - Isolation -                             & Dissemination - COVID19      & 9.32         & 348    & 629  \\ \hline
Fortnight 11 & Trump - USA - Millions - Positive - Genocide                  & Transparency - COVID19       & 5.44         & 203    & 248  \\ \hline
Fortnight 11 & Vaccine - Test - SINOVAC                                      & Prevention - COVID19         & 4.9          & 183    & 267  \\ \hline
Fortnight 11 & Receitas - Effectiveness                                      & Treatment - COVID19          & 1.1          & 41     & 47   \\ \hline
Fortnight 12 & Youth - COVID19 - Virus - Masks - Hydroxychloroquine          & Dissemination - COVID19      & 11.68        & 528    & 995  \\ \hline
Fortnight 12 & Bolsonaro - President - Data - Montage                        & Federal Government - COVID19 & 5.31         & 240    & 330  \\ \hline
Fortnight 12 & Vaccine - Gates - Church                                      & Prevention - COVID19         & 1.66         & 75     & 92   \\ \hline
Fortnight 12 & Federal Supreme Court - Minister - Federal Police             & Politics                     & 1.24         & 56     & 62   \\ \hline
Fortnight 12 & Lacombe - Band - China                                        & International Affairs        & 1.24         & 56     & 65   \\ \hline
\end{tabular}
\caption{ Clusters generated for $@agencialupa$}l\label{tab:agencialupa}
\end{table}

\begin{table}[]
\hspace*{-2.0cm}
\tiny
\begin{tabular}{|l|l|l|c|c|c|}
\hline
Timeframe    & Topics                                                        & Code                         & Cluster (\%) & Vertex & Edge \\ \hline
Fortnight 1  & Bolsonaro - Lula - Temer - Government                                                & Federal Government           & 9.29  & 235 & 337 \\ \hline
Fortnight 1  & Iran - Brazil - Girls                                                                & International Affairs        & 8.82  & 223 & 336 \\ \hline
Fortnight 1  & Sex Education - Damares - Pregnancy - Abstinency                                     & Public Policy                & 3.24  & 82  & 106 \\ \hline
Fortnight 1  & Politics - Ditadura - Sanctions - Complaint                                          & Transparency                 & 2.14  & 54  & 53  \\ \hline
Fortnight 1  & Military - Campaign - Flávio - Protesters                                            & Politics                     & 1.62  & 41  & 42  \\ \hline
Fortnight 2  & Vitamin - Fennel - Wet Throat - Fried Food - Canada - WHO                            & Prevention - COVID19         & 5.52  & 78  & 118 \\ \hline
Fortnight 2  & Virus - Family - Patent - Bat - USA                                                  & Dissemination - COVID19      & 5.02  & 71  & 86  \\ \hline
Fortnight 2  & Bolsonaro - Government - Drop - Occurrences                                          & Federal Government           & 4.67  & 66  & 81  \\ \hline
Fortnight 2  & Data - Information - Law - Security                                                  & Transparency                 & 3.96  & 56  & 75  \\ \hline
Fortnight 2  & Regina Duarte - Military - Actress - Cultural                                        & Media                        & 3.33  & 47  & 57  \\ \hline
Fortnight 3  & Patricia - Bulk messages - Reproductions - Folha de SP                  & Media                        & 7.66  & 137 & 210 \\ \hline
Fortnight 3  & Showers - Mayors - Atypical - Flood - Quality                                        & Public Policy                & 5.54  & 99  & 127 \\ \hline
Fortnight 3  & Sonnante - Vatican - Lula - Pope - USA - Dilma                                       & Politics                     & 3.58  & 64  & 80  \\ \hline
Fortnight 4  & ICMS - Fuels - Gas - Ethanol                                                         & Economy                      & 12.86 & 187 & 314 \\ \hline
Fortnight 4  & Carnival - Millons - SP - Rio - Salvador                                             & Leisure                      & 6.46  & 94  & 137 \\ \hline
Fortnight 4  & Detention - Pre-trial - Brazil - Excess - Prison                                     & Public Security              & 4.75  & 69  & 99  \\ \hline
Fortnight 4  & Vitamin - Vinager - Water - Hot - Lemon                                              & Prevention - COVID19         & 2.96  & 43  & 58  \\ \hline
Fortnight 4  & Bolsonaro - Protest - Artists - Impeachment                                          & Federal Government           & 2.41  & 35  & 48  \\ \hline
Fortnight 5  & Sickness - Symptoms - Cases - Risk - Aglomeration - Minister                         & Dissemination - COVID19      & 10.75 & 208 & 371 \\ \hline
Fortnight 5  & Alcool Gel - Wash - Hands - Vitamin - Vinager                                        & Prevention - COVID19         & 3.57  & 69  & 90  \\ \hline
Fortnight 5  & Forecasts - Growth - GDP - Brazilian                                                 & Economy                      & 2.02  & 39  & 51  \\ \hline
Fortnight 5  & Government - Economy - House - Value - Trump                                         & Politics - COVID19           & 1.6   & 31  & 33  \\ \hline
Fortnight 6  & Data - China - Cases - Death - People                                                & Dissemination - COVID19      & 6.09  & 264 & 517 \\ \hline
Fortnight 6  & Pandemic - Ministers - Health - Mark                                     & Politics - COVID19           & 4.98  & 216 & 330 \\ \hline
Fortnight 6  & Withdrawal - Bolsonaro - Temer - Supermarket - Aid                                   & Economy - COVID19            & 1.55  & 67  & 79  \\ \hline
Fortnight 6  & Social Isolation -  Measures - End - Hygiene                                         & Prevention - COVID19         & 1.18  & 51  & 56  \\ \hline
Fortnight 6  & Drauzio - Varella - Little Flu - COVID19                                             & Treatment - COVID19          & 0.92  & 40  & 42  \\ \hline
Fortnight 7  & Measures - Social Distancing - Heath System - Aglomeration               & Prevention - COVID19         & 7.33  & 185 & 289 \\ \hline
Fortnight 7  & Bolsonaro - President - Mandetta - Osmar Terra                              & Federal Government - COVID19 & 5.11  & 129 & 195 \\ \hline
Fortnight 7  & Emergency Aid - Jobs - Unemployment insurance - Traffic - Flights                    & Economy - COVID19            & 4     & 101 & 130 \\ \hline
Fortnight 8  & Departure - Sergio Moro - Bolsonaro - Federal Police - Resignation                   & Federal Government           & 8.88  & 145 & 214 \\ \hline
Fortnight 8  & Death - COVID19 - States - Health Minitry                                            & Dissemination - COVID19      & 5.21  & 85  & 129 \\ \hline
Fortnight 8  & Isolation - Municipalities - Distancing - Cities                                     & Local - COVID19              & 2.21  & 36  & 46  \\ \hline
Fortnight 9  & Data - Sites - Platform - Transparency - WHO - Registry - Civil                      & Transparency - COVID19       & 8.55  & 88  & 114 \\ \hline
Fortnight 9  & Treatment - Hydroxychloroquine - Antibiotics - Cure - Corticoid                      & Treatment - COVID19          & 8.07  & 83  & 127 \\ \hline
Fortnight 9  & Pandemia - Brazil - Politics - Less                                                  & Media                        & 4.86  & 50  & 58  \\ \hline
Fortnight 9  & Doença - Ceará - Dealth - Minas Gerais                                               & Local - COVID19              & 2.43  & 25  & 27  \\ \hline
Fortnight 10 & Bolsonaro - Statistics - COVID19 - Protest                                           & Dissemination - COVID19      & 10.87 & 169 & 252 \\ \hline
Fortnight 10 & Bolsonaro - Federal Supreme Court - Meeting - Decisions                              & Federal Government           & 8.04  & 125 & 163 \\ \hline
Fortnight 10 & Gaviões da Fiel - Protest - Visited - Politician                                     & Politics                     & 1.67  & 26  & 25  \\ \hline
Fortnight 10 & Trials - Reliable - Clinic - RCTS                                                    & Transparency - COVID19       & 1.61  & 25  & 26  \\ \hline
Fortnight 11 & Investigated - Inquires - Federal Supreme Court - Bolsonaro                          & Federal Government           & 9.44  & 59  & 73  \\ \hline
Fortnight 11 & Government - Company - Foods - Argentina - Fernandez                                 & International Affairs        & 5.92  & 37  & 41  \\ \hline
Fortnight 11 & Ministers - Weintraub - World bank                                                   & Federal Government           & 4.16  & 26  & 28  \\ \hline
Fortnight 11 & COVID19 - Pandemic - Nurse - Entrepreneur - Invoice                                  & Economy - COVID19            & 4     & 25  & 30  \\ \hline
Fortnight 11 & SINOVAC - Contract - Doria - Agreement                                               & Prevention - COVID19         & 3.52  & 22  & 23  \\ \hline
Fortnight 12 & Government - Francisco River - Transposition - PT & Federal Government           & 7.9   & 68  & 82  \\ \hline
Fortnight 12 & Agencia Brasil - EBC                                                                 & Media - COVID19              & 4.41  & 38  & 49  \\ \hline
Fortnight 12 & Federal Supreme Court - Action - Decision - Bolsonaro - Pandemic                     & Federal Government - COVID19 & 4.41  & 38  & 44  \\ \hline
Fortnight 12 & Bolsonaro - Hydroxychloroquine - Ivermectin - Azithromycin - Corticoid               & Treatment - COVID19          & 3.72  & 32  & 48  \\ \hline
\end{tabular}
\caption{ Clusters generated for $@aosfatos$}l\label{tab:aosfatos}
\end{table}

\section{Results}

Our results are divided into three parts, each of them dedicated to answering a research question. The first part is dedicated to present the topics that were covered during the 2020 pandemic and the validity of our computational method (\textbf{RQ1}). In the second part, we analyze the evolution of the political agenda during the pandemic and how it is intertwined with the health crisis(\textbf{RQ2}). Last, we describe the main challenges that arose to identify topics in a period of infodemic (\textbf{RQ3}).

\subsection{Fact-checking shows the evolution of false information}

In order to address \textbf{RQ1}, we present first our findings to the fake news topics that have been debunked or explained by the organization here examined. In response to the COVID-19 crisis and the infodemic, there have been great efforts put into informing the public of the pandemic and its surroundings. The impact of COVID-19 is affecting the way that organizations address immediate and strategic people issues. During the first semester of 2020, Brazilian fact-checking organizations have used social media to combat the spread of false information about the COVID-19 pandemic, which mirrors recently released studies about efforts to tackle misinformation problems via social platforms\cite{Lovari2020SpreadingItaly,Nguyen2020Covid-19Perspective}. Therefore, the structure of our computational model provides a unique framework to interpret the evolution of fact-based information based on the response to the proliferation of misinformation and disinformation online. To validate this, we conduct a brief preliminary round, checking manually random tweets per fortnight.

The clusters obtained using our algorithm are shown in Tables 1 and 2. To illustrate this, we also summarize here the relevant results. False information about the dissemination of the virus, such as pets could transmit COVID-19 and a photo of a wet food market from Indonesia shared as it was in China’s Wuhan were debunked. Also, forms of prevention, such as lemon, orange, fennel, and bicarbonate were spread as forms to protect from coronavirus disease. Another community that our method identified described the rumors and false news about the register of death and over-notification. Also, political topics have also arisen and been developed during this period. Due to the fact that Bolsonaro and his peers disseminated information that was deemed fake or distorted during the pandemic\cite{Statista2020Brazil:COVID-19}, we decided to group them in a category different from ``politics." This was important to show the pandemic has also been influenced by major political and institutional dysfunctions, which will be discussed more in the next section.

\begin{figure}[h!]
	\hspace{-3.0cm}
	\includegraphics[scale=0.18]{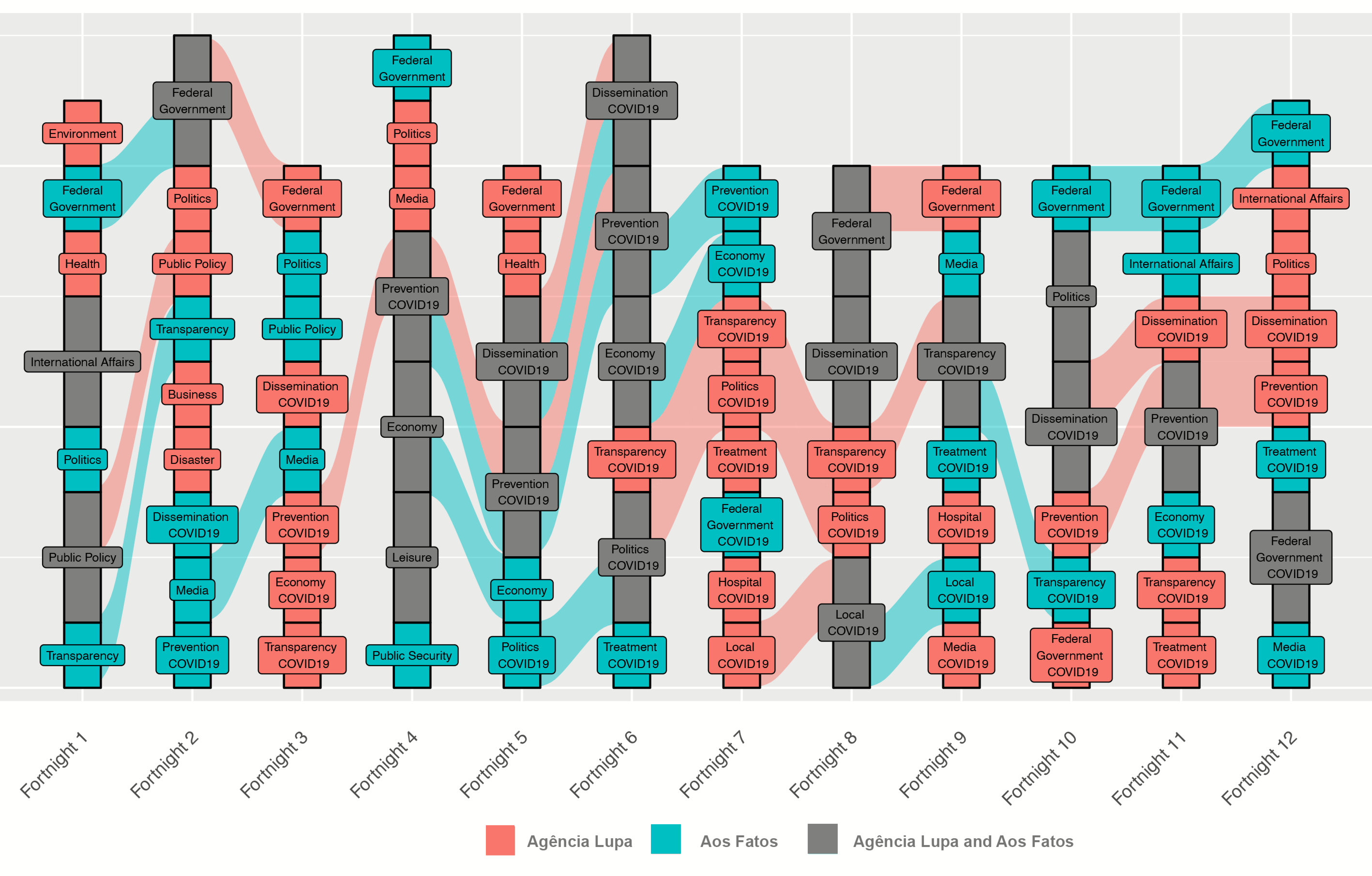}
    \caption{Trends of fake news topics by fortnights (Q1–Q2/2020)}\label{fig:chart} 
\end{figure}

Our algorithm has shown the similarities and differences in the working of fact-checking initiatives to disprove rumors and combat disinformation. A better picture of the evolution of these trends is shown in figure \ref{fig:chart}. In it, the topics are colored according to the organization, being \textit{Aos Fatos} in blue and \textit{Agência Lupa} in red. When a topic is covered by the two organizations at the same time, the gray color is used and is represented by a bar of double size. Furthermore, the flows between one bar and another represent the topics that continued to be debunked in the following month. In this way, we clearly illustrate the continuity of the topics in the time span. 

During fortnight 1, both organizations focused a great part of their work on fact-checking statements about the Australian bushfires and the US-Iran crisis, and also related to public policies, such as the water supply in Rio de Janeiro. Moving to fortnight 2, the Federal Government has gained increasing attention from both organizations. In fortnight 3, our model did not find any major topic that was common to the two initiatives.

From fortnight 4, which includes the two last weeks of February when the first case was reported in Brazil, the discussions about pandemic dominated social media conversations, which provided insight into the fake news spread in online social networks about the COVID-19. The forms of prevention and dissemination were frequently debunked during the whole time considered. These organizations fact-check statements by public figures, major institutions, and other widely circulated claims of interest to society, thus, economy, politics, and Federal Government were also topics that were covered by these two organizations in the same time period. The only exception is during fortnight 4 when Carnival happened, thus, these organizations verified information about leisure activities, such as false information that glitter can damage the health and funding costs for Carnival.

The pandemic crisis is intertwined with uncertainty over the future and a complicated landscape for citizens. This is explained by tapping into topics that cover Federal Government, transparency, and local decisions related to COVID-19 that appeared in our results. There is some fundamental coincidence between these topics in fortnight 12, 9, and 8, respectively. Also, during fortnight 8, politics appeared related to the 2020 pandemic flooded by fake news about supposed empty coffins that were being buried, which induced the population to panic.

A closer look at the literature on fact-checking and fake news argues that the repetition of debunks makes it more believable by the public, however, reveals a number of gaps and shortcomings\cite{Pennycook2018PriorNews,Tandoc2019}. Several fact-checking initiatives around the globe to probe the veracity of news stories in social media but no previous research has investigated if these organizations tend to debunk different stories or are almost exclusively focused on diversifying their content. To better understand the commonalities in the coverage of the different themes that are found in the stories of these two organizations and look for repetition in these topics, we discuss the correlation between the trends that were mapped through their data. 

\begin{figure}[h!]
	%\setlength{\fboxrule}{1pt} 
	%\hspace{-3.0cm}
	\includegraphics[scale=0.6]{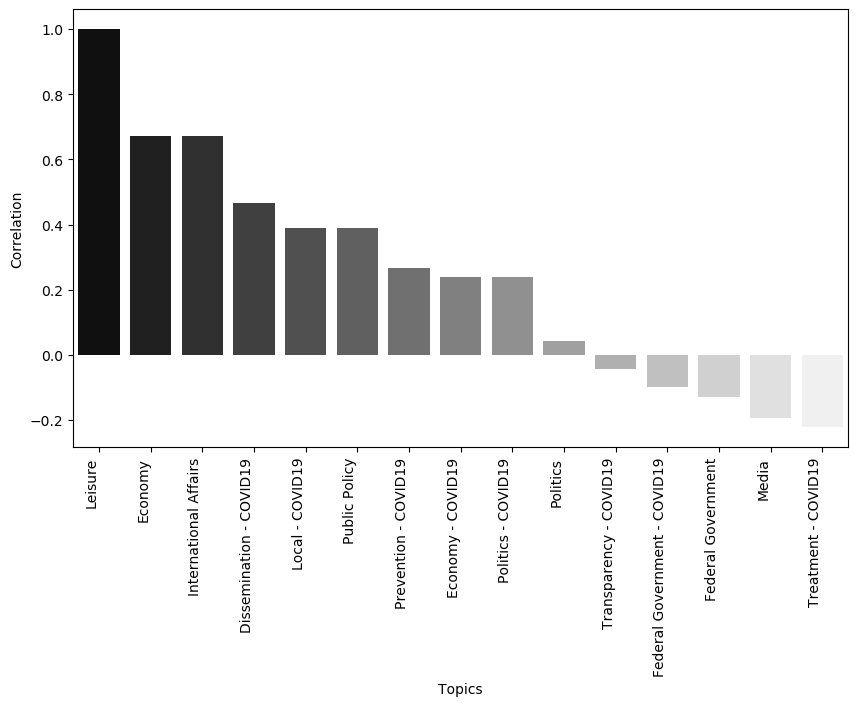}    
    \caption{ Correlation between the topics covered during the pandemic}\label{fig:correlation} 
\end{figure}

Based on the 23 topics found in the time-series analysis of these two fact-checking initiatives, we verified if the same theme emerged in the same time span during the course of the period examined. To this end, a Spearman's correlation, formally referred to as Spearman’s rho, was run to determine the relationship between these topics. Opting for the Spearman correlation is because it is based on the ranked values for each variable rather than the raw data, and also these variables tend to change together, but not necessarily at a constant rate. Also, this correlation is used to assess relationships involving ordinal variables and not linear relationships between two continuous variables, which would require the Pearson correlation. By applying Spearman's correlation on the ranked values based on the time when it was covered, we could find the relationships between the topics not exclusively when it happens but also the lack of this topic in a certain period of time, which are presented in figure \ref{fig:correlation}.

Figure \ref{fig:correlation} shows only one topic has a strong correlation with the time-series data. Leisure was common to both fact-checking initiatives in the same period of time due to the period of the most important event in the country, the Brazilian Carnival, which happened in the last fortnight of February. The second strongest correlation is the topic Economy due to the financial crisis that the country is going through and the polemics involving president Jair Bolsonaro that have a strong impact on the economy. And it was also the theme of the third strongest correlation but in this case related to misleading information about the impact of COVID-19 in the economy. Overall, our studies indicate that there are overlapping in the coverage areas of these two organizations. However, in the major part, these initiatives pick and choose what to scrutinize, limiting their main points of incidence. Also, a few topics last for a very long time, which does not echo studies that show that the message repetition made it more believable\cite{Pennycook2018PriorNews,Tandoc2019}.

\subsection{Political agenda is used to draw the attention away from the health crisis}

Our algorithm demystifies that fact-checking organizations were only overwhelmed by coronavirus misinformation (\textbf{RQ2}). Understanding this phenomenon is crucial, given the short and long-term effects of fake news not only on individuals and organizations but also on societies and cultures. These organizations debunk politics and government power and offer a hopeful outline of a world both free and more democratic, as shown in the literature\cite{Tandoc2019,Walter2019NewsApproach,Graves2018Aa}. When politicians fail on how to translate the scientific evidence into policy and action, it amplifies the influence of rumors, informal news sources, and fringe journalism. This raises the importance of understanding change over time in relation to each case, rather than relying only on a comparison of volumes, and we turn to figure \ref{fig:chart} for this.

Furthermore, by uncovering false and misleading information, these fact-checking organizations shed light on the influence of politics on the disproportionate rates of illness and death experienced during the COVID-19 situation. A first observation that highlights the imbalance between fortnights is significant. The evolution in time shows a continuously growing trend towards misinformation about the dissemination and prevention of COVID-19 but also about themes related to the Federal Government that did not arise from the pandemic. While in the fortnights 6 and 7 only topics related to COVID-19 were fact-checked, the following weeks were composed by exposure of stories that are in the public interest corresponding to mainly Federal Government but also media, politics, and international affairs but not related to health crises.

This echoes with the political situation in Brazil that helps to explain these developments. In the due course, the country had three ministers who resigned, being two of them Health Ministries. Also, in June, the government had even stopped publishing total numbers of deaths and cases\cite{Phillips2020BrazilSite}. This shows how disinformation thrives with political instability and conflict.

\begin{figure}[h!]
	\hspace{-2cm}
	\includegraphics[scale=0.55]{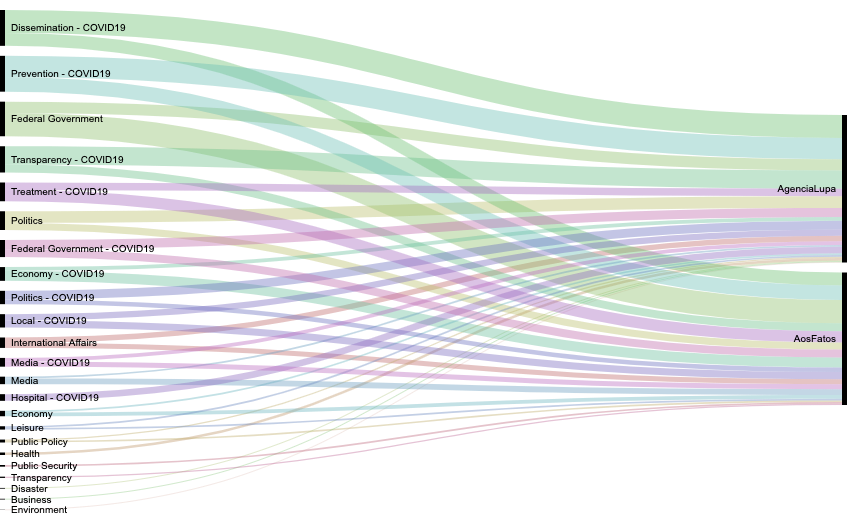}    
    \caption{Topics covered by these two organizations during the 2020 pandemic crisis}\label{fig:topics} 

\end{figure}

Our main approach was to compare relative volumes of different classes by the organization as shown in figure \ref{fig:topics}. This illustration shows the topics exposed by the falseness of claims, dissemination, and prevention have led, followed next by the Federal Government. Between the two fact-checking outlets, \textit{Aos Fatos} have checked more claims related to the government and the president Bolsonaro than \textit{Agência Lupa}. This demonstrates how the silence on the influence of politics and also the shifting focus from deadly virus to politics, law, and jobs is a matter of serious concern. An example is a fake news story about the former Brazil's justice minister, Sergio Moro, when he resigned: a rumor was spread in social media that citing that the former minister prevented an investigation of other politicians when he was a judge to tarnish his reputation. Similarly, other events happened that have shifted the focus away from the pandemic.

Beyond a computational method, we propose in this article to investigate the interplay between fake news, fact-checking, politics, and the ongoing health crisis, and to outline the temporal evolution of debunked topics. As shown in these results, this is particularly important in Brazil, once we show that the political agenda has gained more attention than the health crisis due to the political actors and institutions in the country. Thus, our finding highlights that there is a distinct number of topics that were specifically focused on politics and COVID-19, while others were exclusively about politics. Consequently, it focuses people off the health crisis and keeps going with the disinformation agenda about politics. A new approach is therefore needed to understand how fact-checking organizations decide which topics to debunk and the political campaigns that are currently fueling social media platforms with fake news.

\subsection{The infodemic: a very specific scenario}
In our work about fact-checking political claims and debunking viral deceptions, we have found a tremendous amount of misinformation on the coronavirus pandemic. This was expected as the unprecedented COVID-19 crisis triggered an avalanche of unprecedented challenges both for people and society. In this context, the restriction to a very specific scenario and model hinders interoperability with other components. Our data shows that the majority of topics covered are based on the dissemination and prevention of the COVID-19 outbreak. Also, fake news about transparency and treatment dominates the hot topics that were debunked.

These potential contradictions that are created by fake news generate confusion and distrust in society and it can undermine an effective public health response. This fact also deals with the limitation that infodemic brought by spreading an excessive amount of information about an issue, which makes it confusing and hinders to identify a solution (\textbf{RQ3}). Therefore, this is fertile ground for the dissemination of fake narratives as we identified by the number of refuted or checked claims by fact-checkers.

Furthermore, our data has shown that some topics lasted longer than others. Before the first case (since fortnight 2), the fact-checking about dissemination and prevention have been checked by at least one of the organizations here studied. Although this period has an information overload about the same theme, this asymmetry does not influence our investigation. The clusters were generated even with an insufficient lexicon coverage, as described by the percentage of community in relation to the whole dataset in Tables \ref{tab:agencialupa} and \ref{tab:agencialupa}. 

These examples show that the use of a computational method based on Markov chains is more effective than other traditional methods for topic modeling, such as LDA and LSA, which require more lexicon and more corpus. Our study also provides an opportunity for the examination of relationships between relative volumes and temporal dynamics, whether across a whole dataset or for individual Twitter accounts.

\section{Discussion and Conclusion}

Fake news has existed, still exists, and will continue to exist. The digital age and the rapid expansion of social media have provided a breeding ground for the dissemination of misleading and false information. Thus, fake news became a global phenomenon with no solution on the horizon. Debunks became a stopgap measure to reduce belief in false claims or prevent the uptake of misinformation. In fact, the importance of fact-checking initiatives lies in their ability to uncover and debunk hoaxes on social media platforms to facilitate the exposure of stories that are in the public interest. These fact-checks are used as forms to mitigate the spread of harmful misinformation. Collectively, our results provide a potential mechanism for identifying the fake news topics that arose during the time span considered.

The findings presented here provide us with first insights into fake news agenda based on the fact-checking content posted by specialized organizations on Twitter. Against the wave of speculations and fake news\cite{Bruns2020CoronaFacebook}, scientific facts matter. The infodemic has shown presence and allowed it to escalate to a level that requires a coordinated response. It has triggered an army of fact-checkers and debunkers worldwide to verify these false or misleading statements. In Brazil, the fact-checking agencies \textit{Lupa} and \textit{Aos Fatos} showed the relevance of fact-checking in a situation where political decisions remain incongruent to the global health crises. Their debunks covered not only this infodemic but also a range of topics that were necessary to address in autocratic populism promoted by Jair Bolsonaro.

Our method has shown the widespread potential to scale up this analysis from two accounts with 5115 tweets to a broad number of accounts or tweets. In our analysis of the two major fact-checking organizations in Brazil, we have found the similarities and differences in what was debunked by these organizations, which shows a different perspective from the existing literature stating that repetition makes fact-checking more believable \cite{Pennycook2018PriorNews,Tandoc2019}. Only a few topics lasted more than one fortnight. Also, this applied to the topics that were covered during the fortnights and were not continued into the following periods. 

The analysis leads to the following conclusions: these organizations do not necessarily debunk the same topic and also the fact-checking agenda does not necessarily reflect the situation that currently exists within a country. Furthermore, the literature pertaining to fact-checking practices strongly suggests that a detailed debunking message when presented to the user repeatedly in varying time intervals correlated positively with the debunking effect \cite{Chan2017Debunking:Misinformation,Pennycook2018PriorNews,Tandoc2019}. However, in our study, the fact-checking agencies adopted to cover distinct themes. 

This article contributes to a growing corpus of research of fake news by developing a method that can determine the main characteristic of the fake news and fact-checking genres that can help news organizations to identify topics that rose or fell in popularity in the social media platforms and look at unique ways and angles to tackle this issue. Also, identifying categories that experience sharp temporal increases during the pandemic is important to set an agenda and include support for efforts to resolve these issues and identify means of fostering cooperation between different fact-checking initiatives.

The analysis presented here also suggests that the political agenda has shifted away from the focus of the health crisis and gave more attention to the political actors and institutions. An extreme right-wing political agenda is used to legitimize the leader’s actions and remove the focus from the crisis. Consequently, politics and pandemic have been intertwined in varying degrees, which causes continuing difficulties for tackling misinformation. Future studies should aim to replicate results in a larger set of time to an integrated overview of pandemic and its evolution over time. Also, our computational method can be applied to political entities to cluster what they are posting on social media during these health crises. In addition, future research might automate the time spans to optimize the division of topics per period.

Although this study is very beneficial to the field, limitations should be addressed. Unfortunately, there are not enough tweets from other fact-checking initiatives. We, therefore, concentrate on two major fact-checking organizations, and we do not investigate the numerous other smaller initiatives existing in the platform because the number of tweets produced by them during the period was not enough to be considered in this study. This can be a potential limitation of this observational cohort study. Future studies should aim to replicate results in a more substantial dataset, including more organizations and countries. Also, a different time interval can help to include more fact-checking initiatives that have less number of tweets.

Last, a critical open question is whether this method is generic and can be applied not only to tweets but to a different range of datasets with distinct sizes. This experiment adds to a growing corpus of research showing that even with limited text corpora, we were able to cluster topics. This result shows that our method has the potential to be applied to other contexts, but future studies could fruitfully explore this issue.

Fact-checking is an effort to tackle misinformation problems but there is no one-size-fits-all solution to fake news. But in a period of uncertainty and almost constantly changing news and new research is added regularly while much has not yet been discovered, fact-checking provides a service by clarifying the assessment of a claim by delivering context and background information in a situation that any little bit makes a difference. The COVID-19 pandemic is just a reminder about the huge problem of misinformation, disinformation, and hoaxes that have become part of our lives now.

\section{Acknowledgements}

This project was partially funded by the European Union’s Horizon 2020 research and innovation programme under the Marie Sklodowska Curie grant agreement No 765140.

% \section*{References}


\begin{thebibliography}{10}
\expandafter\ifx\csname url\endcsname\relax
  \def\url#1{\texttt{#1}}\fi
\expandafter\ifx\csname urlprefix\endcsname\relax\def\urlprefix{URL }\fi
\expandafter\ifx\csname href\endcsname\relax
  \def\href#1#2{#2} \def\path#1{#1}\fi

\bibitem{Carpanez2018VejaGuaruja}
J.~Carpanez,
  \href{https://www1.folha.uol.com.br/cotidiano/2018/09/veja-o-passo-a-passo-da-noticia-falsa-que-acabou-em-tragedia-em-guaruja.shtml}{{Veja
  o passo a passo da not{\'{i}}cia falsa que acabou em trag{\'{e}}dia em
  Guaruj{\'{a}}}} (9 2018).
\newline\urlprefix\url{https://www1.folha.uol.com.br/cotidiano/2018/09/veja-o-passo-a-passo-da-noticia-falsa-que-acabou-em-tragedia-em-guaruja.shtml}

\bibitem{Schapals2018FakeFacts}
A.~K. Schapals, \href{https://doi.org/10.1080/17512786.2018.1511822
  https://linkinghub.elsevier.com/retrieve/pii/S1062940817302656
  https://www.tandfonline.com/doi/full/10.1080/17512786.2018.1511822}{{Fake
  News: Australian and British journalists’ role perceptions in an era of
  “alternative facts”}}, Journalism Practice 12~(8) (2018) 976--985.
\newblock \href {http://dx.doi.org/10.1080/17512786.2018.1511822}
  {\path{doi:10.1080/17512786.2018.1511822}}.
\newline\urlprefix\url{https://doi.org/10.1080/17512786.2018.1511822
  https://linkinghub.elsevier.com/retrieve/pii/S1062940817302656
  https://www.tandfonline.com/doi/full/10.1080/17512786.2018.1511822}

\bibitem{Lazer2018}
D.~M. Lazer, M.~A. Baum, Y.~Benkler, A.~J. Berinsky, K.~M. Greenhill,
  F.~Menczer, M.~J. Metzger, B.~Nyhan, G.~Pennycook, D.~Rothschild,
  M.~Schudson, S.~A. Sloman, C.~R. Sunstein, E.~A. Thorson, D.~J. Watts, J.~L.
  Zittrain, {The science of fake news: Addressing fake news requires a
  multidisciplinary effort}, Science 359~(6380) (2018) 1094--1096.
\newblock \href {http://dx.doi.org/10.1126/science.aao2998}
  {\path{doi:10.1126/science.aao2998}}.

\bibitem{Tandoc2018DefiningDefinitions}
E.~C. Tandoc, Z.~W. Lim, R.~Ling,
  \href{https://www.tandfonline.com/action/journalInformation?journalCode=rdij20
  https://www.tandfonline.com/doi/full/10.1080/21670811.2017.1360143}{{Defining
  “Fake News”: A typology of scholarly definitions}}, Digital Journalism
  6~(2) (2018) 137--153.
\newblock \href {http://dx.doi.org/10.1080/21670811.2017.1360143}
  {\path{doi:10.1080/21670811.2017.1360143}}.
\newline\urlprefix\url{https://www.tandfonline.com/action/journalInformation?journalCode=rdij20
  https://www.tandfonline.com/doi/full/10.1080/21670811.2017.1360143}

\bibitem{WHO2020}
{WHO},
  \href{https://www.un.org/en/un-coronavirus-communications-team/un-tackling-‘infodemic’-misinformation-and-cybercrime-covid-19}{{UN
  tackles ‘infodemic’ of misinformation and cybercrime in COVID-19 crisis}}
  (2020).
\newline\urlprefix\url{https://www.un.org/en/un-coronavirus-communications-team/un-tackling-‘infodemic’-misinformation-and-cybercrime-covid-19}

\bibitem{Tandoc2019}
E.~C. Tandoc, \href{https://doi.org/10.1111/soc4.12724}{{The facts of fake
  news: A research review}}, Sociology Compass 13~(9) (2019) e12724.
\newblock \href {http://dx.doi.org/10.1111/soc4.12724}
  {\path{doi:10.1111/soc4.12724}}.
\newline\urlprefix\url{https://doi.org/10.1111/soc4.12724}

\bibitem{Tambuscio2018NetworkFact-checking}
M.~Tambuscio, D.~F.~M. Oliveira, G.~L. Ciampaglia, G.~Ruffo,
  \href{https://doi.org/10.1007/s42001-018-0018-9}{{Network segregation in a
  model of misinformation and fact-checking}}, Journal of Computational Social
  Science 1~(2) (2018) 261--275.
\newblock \href {http://dx.doi.org/10.1007/s42001-018-0018-9}
  {\path{doi:10.1007/s42001-018-0018-9}}.
\newline\urlprefix\url{https://doi.org/10.1007/s42001-018-0018-9}

\bibitem{Dobbs2012}
M.~Dobbs,
  \href{http://newamerica.net/sites/newamerica.net/files/policydocs/The_Rise_of_Political_Fact-checking.pdf.pdf}{{The
  rise of political fact-checking: How Reagan inspired a journalistic movement:
  A reporter's eye view}}, Tech. rep., New America Foundation (2 2012).
\newline\urlprefix\url{http://newamerica.net/sites/newamerica.net/files/policydocs/The_Rise_of_Political_Fact-checking.pdf.pdf}

\bibitem{Walter2019NewsApproach}
D.~Walter, Y.~Ophir,
  \href{https://www.tandfonline.com/action/journalInformation?journalCode=hcms20}{{News
  Frame Analysis: An Inductive Mixed-method Computational Approach}},
  Communication Methods and Measures 13~(4) (2019) 248--266.
\newblock \href {http://dx.doi.org/10.1080/19312458.2019.1639145}
  {\path{doi:10.1080/19312458.2019.1639145}}.
\newline\urlprefix\url{https://www.tandfonline.com/action/journalInformation?journalCode=hcms20}

\bibitem{Flew2019}
T.~Flew, P.~Iosifidis,
  \href{http://journals.sagepub.com/doi/10.1177/1748048519880721}{{Populism,
  globalisation and social media}}, International Communication Gazette 82~(1)
  (2020) 7--25.
\newblock \href {http://dx.doi.org/10.1177/1748048519880721}
  {\path{doi:10.1177/1748048519880721}}.
\newline\urlprefix\url{http://journals.sagepub.com/doi/10.1177/1748048519880721}

\bibitem{Goovaerts2020}
I.~Goovaerts, S.~Marien,
  \href{https://www.tandfonline.com/action/journalInformation?journalCode=upcp20}{{Uncivil
  Communication and Simplistic Argumentation: Decreasing Political Trust,
  Increasing Persuasive Power?}}, Political Communication\href
  {http://dx.doi.org/10.1080/10584609.2020.1753868}
  {\path{doi:10.1080/10584609.2020.1753868}}.
\newline\urlprefix\url{https://www.tandfonline.com/action/journalInformation?journalCode=upcp20}

\bibitem{Walter2020}
N.~Walter, J.~Cohen, R.~L. Holbert, Y.~Morag,
  \href{https://www.tandfonline.com/action/journalInformation?journalCode=upcp20
  https://www.tandfonline.com/doi/full/10.1080/10584609.2019.1668894}{{Fact-Checking:
  A Meta-Analysis of What Works and for Whom}}, Political Communication 37~(3)
  (2020) 350--375.
\newblock \href {http://dx.doi.org/10.1080/10584609.2019.1668894}
  {\path{doi:10.1080/10584609.2019.1668894}}.
\newline\urlprefix\url{https://www.tandfonline.com/action/journalInformation?journalCode=upcp20
  https://www.tandfonline.com/doi/full/10.1080/10584609.2019.1668894}

\bibitem{Tandoc2020WhenOutbreak}
E.~C. Tandoc, J.~C.~B. Lee,
  \href{http://journals.sagepub.com/doi/10.1177/1461444820968212}{{When viruses
  and misinformation spread: How young Singaporeans navigated uncertainty in
  the early stages of the COVID-19 outbreak}}, New Media and Society (2020)
  146144482096821\href {http://dx.doi.org/10.1177/1461444820968212}
  {\path{doi:10.1177/1461444820968212}}.
\newline\urlprefix\url{http://journals.sagepub.com/doi/10.1177/1461444820968212}

\bibitem{Bakir2018FakeSolutions}
V.~Bakir, A.~McStay,
  \href{https://www.tandfonline.com/action/journalInformation?journalCode=rdij20
  https://www.tandfonline.com/doi/full/10.1080/21670811.2017.1345645}{{Fake
  News and The Economy of Emotions: Problems, causes, solutions}}, Digital
  Journalism 6~(2) (2018) 154--175.
\newblock \href {http://dx.doi.org/10.1080/21670811.2017.1345645}
  {\path{doi:10.1080/21670811.2017.1345645}}.
\newline\urlprefix\url{https://www.tandfonline.com/action/journalInformation?journalCode=rdij20
  https://www.tandfonline.com/doi/full/10.1080/21670811.2017.1345645}

\bibitem{Braun2019FakeJournalism}
J.~A. Braun, J.~L. Eklund,
  \href{https://www.tandfonline.com/doi/full/10.1080/21670811.2018.1556314}{{Fake
  News, Real Money: Ad Tech Platforms, Profit-Driven Hoaxes, and the Business
  of Journalism}}, Digital Journalism 0~(0) (2019) 1--21.
\newblock \href {http://dx.doi.org/10.1080/21670811.2018.1556314}
  {\path{doi:10.1080/21670811.2018.1556314}}.
\newline\urlprefix\url{https://www.tandfonline.com/doi/full/10.1080/21670811.2018.1556314}

\bibitem{Graves2018UnderstandingFact-Checking}
L.~Graves, \href{https://www.poynter.org/news/review-live-fact-checking-
  http://foreignpolicy.com/2016/05/06/understanding-the-promise-and-limits-of-sanctions/}{{Understanding
  the Promise and Limits of Automated Fact-Checking}}, Tech. Rep. February,
  Reuters Institute, Oxford (2018).
\newblock \href {http://dx.doi.org/10.1007/s13398-014-0173-7.2}
  {\path{doi:10.1007/s13398-014-0173-7.2}}.
\newline\urlprefix\url{https://www.poynter.org/news/review-live-fact-checking-
  http://foreignpolicy.com/2016/05/06/understanding-the-promise-and-limits-of-sanctions/}

\bibitem{Allcott2019}
H.~Allcott, M.~Gentzkow, C.~Yu,
  \href{http://journals.sagepub.com/doi/10.1177/2053168019848554}{{Trends in
  the diffusion of misinformation on social media}}, Research and Politics
  6~(2) (2019) 205316801984855.
\newblock \href {http://dx.doi.org/10.1177/2053168019848554}
  {\path{doi:10.1177/2053168019848554}}.
\newline\urlprefix\url{http://journals.sagepub.com/doi/10.1177/2053168019848554}

\bibitem{Prasetya2020AMedia}
H.~A. Prasetya, T.~Murata,
  \href{https://link.springer.com/articles/10.1186/s40649-019-0076-z
  https://link.springer.com/article/10.1186/s40649-019-0076-z}{{A model of
  opinion and propagation structure polarization in social media}},
  Computational Social Networks 7~(1) (2020) 1--35.
\newblock \href {http://dx.doi.org/10.1186/s40649-019-0076-z}
  {\path{doi:10.1186/s40649-019-0076-z}}.
\newline\urlprefix\url{https://link.springer.com/articles/10.1186/s40649-019-0076-z
  https://link.springer.com/article/10.1186/s40649-019-0076-z}

\bibitem{Nyhan2010WhenMisperceptions}
B.~Nyhan, J.~Reifler,
  \href{https://link.springer.com/article/10.1007/s11109-010-9112-2}{{When
  corrections fail: The persistence of political misperceptions}}, Political
  Behavior 32~(2) (2010) 303--330.
\newblock \href {http://dx.doi.org/10.1007/s11109-010-9112-2}
  {\path{doi:10.1007/s11109-010-9112-2}}.
\newline\urlprefix\url{https://link.springer.com/article/10.1007/s11109-010-9112-2}

\bibitem{Wood2018TheAdherence}
T.~Wood, E.~Porter, {The Elusive Backfire Effect: Mass Attitudes’ Steadfast
  Factual Adherence}, Political Behavior 41.
\newblock \href {http://dx.doi.org/10.1007/s11109-018-9443-y}
  {\path{doi:10.1007/s11109-018-9443-y}}.

\bibitem{Ecker2020TheFact-checks}
U.~K.~H. Ecker, Z.~O'Reilly, J.~S. Reid, E.~P. Chang,
  \href{https://doi.org/10.1111/bjop.12383}{{The effectiveness of short-format
  refutational fact-checks}}, British Journal of Psychology 111~(1) (2020)
  36--54.
\newblock \href {http://dx.doi.org/https://doi.org/10.1111/bjop.12383}
  {\path{doi:https://doi.org/10.1111/bjop.12383}}.
\newline\urlprefix\url{https://doi.org/10.1111/bjop.12383}

\bibitem{Ciampaglia2018FightingMisinformation}
G.~L. Ciampaglia, \href{https://doi.org/10.1007/s42001-017-0005-6}{{Fighting
  fake news: a role for computational social science in the fight against
  digital misinformation}}, Journal of Computational Social Science 1~(1)
  (2018) 147--153.
\newblock \href {http://dx.doi.org/10.1007/s42001-017-0005-6}
  {\path{doi:10.1007/s42001-017-0005-6}}.
\newline\urlprefix\url{https://doi.org/10.1007/s42001-017-0005-6}

\bibitem{Graves2018Aa}
L.~Graves,
  \href{https://www.tandfonline.com/action/journalInformation?journalCode=rjos20}{{Boundaries
  Not Drawn: Mapping the institutional roots of the global fact-checking
  movement}}, Journalism Studies 19~(5) (2018) 613--631.
\newblock \href {http://dx.doi.org/10.1080/1461670X.2016.1196602}
  {\path{doi:10.1080/1461670X.2016.1196602}}.
\newline\urlprefix\url{https://www.tandfonline.com/action/journalInformation?journalCode=rjos20}

\bibitem{Shao2016}
C.~Shao, G.~L. Ciampaglia, A.~Flammini, F.~Menczer,
  \href{http://dl.acm.org/citation.cfm?doid=2872518.2890098}{{Hoaxy}}, in:
  Proceedings of the 25th International Conference Companion on World Wide Web
  - WWW '16 Companion, ACM Press, New York, New York, USA, 2016, pp. 745--750.
\newblock \href {http://dx.doi.org/10.1145/2872518.2890098}
  {\path{doi:10.1145/2872518.2890098}}.
\newline\urlprefix\url{http://dl.acm.org/citation.cfm?doid=2872518.2890098}

\bibitem{Coddington2014}
M.~Coddington, L.~Molyneux, R.~G. Lawrence,
  \href{http://journals.sagepub.com/doi/10.1177/1940161214540942}{{Fact
  Checking the Campaign: How Political Reporters Use Twitter to Set the Record
  Straight (or Not)}}, International Journal of Press/Politics 19~(4) (2014)
  391--409.
\newblock \href {http://dx.doi.org/10.1177/1940161214540942}
  {\path{doi:10.1177/1940161214540942}}.
\newline\urlprefix\url{http://journals.sagepub.com/doi/10.1177/1940161214540942}

\bibitem{Vosoughi2018a}
S.~Vosoughi, D.~Roy, S.~Aral,
  \href{https://www.sciencemag.org/lookup/doi/10.1126/science.aap9559}{{The
  spread of true and false news online}}, Science 359~(6380) (2018) 1146--1151.
\newblock \href {http://dx.doi.org/10.1126/science.aap9559}
  {\path{doi:10.1126/science.aap9559}}.
\newline\urlprefix\url{https://www.sciencemag.org/lookup/doi/10.1126/science.aap9559}

\bibitem{Vargo2018}
C.~J. Vargo, L.~Guo, M.~A. Amazeen,
  \href{https://doi.org/10.1177/1461444817712086
  http://www.ncbi.nlm.nih.gov/pubmed/19204204
  http://www.pubmedcentral.nih.gov/articlerender.fcgi?artid=PMC2667820
  http://journals.sagepub.com/doi/10.1177/1461444817712086}{{The agenda-setting
  power of fake news: A big data analysis of the online media landscape from
  2014 to 2016}}, New Media and Society 20~(5) (2018) 2028--2049.
\newblock \href {http://dx.doi.org/10.1177/1461444817712086}
  {\path{doi:10.1177/1461444817712086}}.
\newline\urlprefix\url{https://doi.org/10.1177/1461444817712086
  http://www.ncbi.nlm.nih.gov/pubmed/19204204
  http://www.pubmedcentral.nih.gov/articlerender.fcgi?artid=PMC2667820
  http://journals.sagepub.com/doi/10.1177/1461444817712086}

\bibitem{Miller2015}
P.~R. Miller, P.~J. Conover,
  \href{http://journals.sagepub.com/doi/10.1177/1065912915577208}{{Red and Blue
  States of Mind: Partisan Hostility and Voting in the United States}},
  Political Research Quarterly 68~(2) (2015) 225--239.
\newblock \href {http://dx.doi.org/10.1177/1065912915577208}
  {\path{doi:10.1177/1065912915577208}}.
\newline\urlprefix\url{http://journals.sagepub.com/doi/10.1177/1065912915577208}

\bibitem{Matos2020}
F.~J.~R. Paumgartten, I.~F. Delgado, L.~D.~R. Pitta, A.~C. A. X.~d. Oliveira,
  \href{https://doi.org/10.22239/2317-269x.01596
  http://www.visaemdebate.incqs.fiocruz.br/index.php/visaemdebate/article/view/1596/1152}{{Drug
  repurposing clinical trials in the search for life-saving COVID-19 therapies;
  research targets and methodological and ethical issues}}, Vigil{\^{a}}ncia
  Sanit{\'{a}}ria em Debate 8~(2) (2020) 39--53.
\newblock \href {http://dx.doi.org/10.22239/2317-269x.01596}
  {\path{doi:10.22239/2317-269x.01596}}.
\newline\urlprefix\url{https://doi.org/10.22239/2317-269x.01596
  http://www.visaemdebate.incqs.fiocruz.br/index.php/visaemdebate/article/view/1596/1152}

\bibitem{Boyd2012}
D.~Boyd, K.~Crawford,
  \href{https://www.tandfonline.com/doi/abs/10.1080/1369118X.2012.678878
  http://www.tandfonline.com/doi/abs/10.1080/1369118X.2012.678878
  https://www.tandfonline.com/action/journalInformation?journalCode=rics20}{{Critical
  questions for big data: Provocations for a cultural, technological, and
  scholarly phenomenon}}, Information Communication and Society 15~(5) (2012)
  662--679.
\newblock \href {http://dx.doi.org/10.1080/1369118X.2012.678878}
  {\path{doi:10.1080/1369118X.2012.678878}}.
\newline\urlprefix\url{https://www.tandfonline.com/doi/abs/10.1080/1369118X.2012.678878
  http://www.tandfonline.com/doi/abs/10.1080/1369118X.2012.678878
  https://www.tandfonline.com/action/journalInformation?journalCode=rics20}

\bibitem{Stieglitz2014SocialAnalytics}
S.~Stieglitz, L.~Dang-Xuan, A.~Bruns, C.~Neuberger,
  \href{http://link.springer.com/10.1007/s11576-014-0407-5}{{Social media
  analytics}}, Business and Information Systems Engineering 6~(2) (2014)
  101--109.
\newblock \href {http://dx.doi.org/10.1007/s11576-014-0407-5}
  {\path{doi:10.1007/s11576-014-0407-5}}.
\newline\urlprefix\url{http://link.springer.com/10.1007/s11576-014-0407-5}

\bibitem{Recuero2004TEORIAFotologs}
R.~Recuero, {TEORIA DAS REDES E REDES SOCIAIS NA INTERNET: Considera{\c{c}}ões
  sobre o Orkut, os Weblogs e os Fotologs}, in: IV Encontro dos N{\'{u}}cleos
  de Pesquisa da XXVII INTERCOM, Porto Alegre, 2004.

\bibitem{Gonzalez-Bailon2016NetworkedMedia}
S.~Gonz{\'{a}}lez-Bail{\'{o}}n, N.~Wang, {Networked discontent: The anatomy of
  protest campaigns in social media}, Social Networks 44 (2016) 95--104.
\newblock \href {http://dx.doi.org/10.1016/j.socnet.2015.07.003}
  {\path{doi:10.1016/j.socnet.2015.07.003}}.

\bibitem{Pang2009OpinionAnalysis}
B.~Pang, L.~Lee, {Opinion mining and sentiment analysis}, Computational
  Linguistics 35~(2) (2009) 311--312.
\newblock \href {http://dx.doi.org/10.1162/coli.2009.35.2.311}
  {\path{doi:10.1162/coli.2009.35.2.311}}.

\bibitem{Aston2014TwitterPerceptron}
N.~Aston, J.~Liddle, W.~Hu, \href{http://www.scirp.org/journal/jcc
  http://dx.doi.org/10.4236/jcc.2014.23002}{{Twitter Sentiment in Data Streams
  with Perceptron}}, Journal of Computer and Communications 02~(03) (2014)
  11--16.
\newblock \href {http://dx.doi.org/10.4236/jcc.2014.23002}
  {\path{doi:10.4236/jcc.2014.23002}}.
\newline\urlprefix\url{http://www.scirp.org/journal/jcc
  http://dx.doi.org/10.4236/jcc.2014.23002}

\bibitem{Davidov2010EnhancedSmileys}
D.~Davidov, O.~Tsur, A.~Rappoport, {Enhanced sentiment learning using twitter
  hashtags and smileys}, Tech. rep. (2010).

\bibitem{Bruns2020CoronaFacebook}
A.~Bruns, S.~Harrington, E.~Hurcombe,
  \href{http://journals.sagepub.com/doi/10.1177/1329878X20946113}{{‘Corona?
  5G? or both?’: the dynamics of COVID-19/5G conspiracy theories on
  Facebook}}, Media International Australia (2020) 1329878X2094611\href
  {http://dx.doi.org/10.1177/1329878X20946113}
  {\path{doi:10.1177/1329878X20946113}}.
\newline\urlprefix\url{http://journals.sagepub.com/doi/10.1177/1329878X20946113}

\bibitem{Gruzd2020GoingTwitter}
A.~Gruzd, P.~Mai,
  \href{http://journals.sagepub.com/doi/10.1177/2053951720938405}{{Going viral:
  How a single tweet spawned a COVID-19 conspiracy theory on Twitter}}, Big
  Data {\&} Society 7~(2) (2020) 205395172093840.
\newblock \href {http://dx.doi.org/10.1177/2053951720938405}
  {\path{doi:10.1177/2053951720938405}}.
\newline\urlprefix\url{http://journals.sagepub.com/doi/10.1177/2053951720938405}

\bibitem{Karami2020}
A.~Karami, M.~Lundy, F.~Webb, Y.~K. Dwivedi,
  \href{https://ieeexplore.ieee.org/document/9047963/}{{Twitter and Research: A
  Systematic Literature Review Through Text Mining}}, IEEE Access 8 (2020)
  67698--67717.
\newblock \href {http://dx.doi.org/10.1109/ACCESS.2020.2983656}
  {\path{doi:10.1109/ACCESS.2020.2983656}}.
\newline\urlprefix\url{https://ieeexplore.ieee.org/document/9047963/}

\bibitem{Lu2012NumericalBodies}
H.~Lu, C.~Yang, R.~L{\"{o}}hner, \href{https://doi.org/}{{Numerical Studies of
  Green Water Impact On Fixed And Moving Bodies}}, International Journal of
  Offshore and Polar Engineering 22~(02) (2012) 10.
\newline\urlprefix\url{https://doi.org/}

\bibitem{Keller2020NewsTopics}
T.~R. Keller, V.~Hase, J.~Thaker, D.~Mahl, M.~S. Sch{\"{a}}fer,
  \href{https://www.tandfonline.com/action/journalInformation?journalCode=renc20}{{News
  Media Coverage of Climate Change in India 1997–2016: Using Automated
  Content Analysis to Assess Themes and Topics}}, Environmental Communication
  14~(2) (2020) 219--235.
\newblock \href {http://dx.doi.org/10.1080/17524032.2019.1643383}
  {\path{doi:10.1080/17524032.2019.1643383}}.
\newline\urlprefix\url{https://www.tandfonline.com/action/journalInformation?journalCode=renc20}

\bibitem{Zhao2011ComparingModels}
W.~X. Zhao, J.~Jiang, J.~Weng, J.~He, E.~P. Lim, H.~Yan, X.~Li,
  \href{http://query.nytimes.com/search/}{{Comparing twitter and traditional
  media using topic models}}, in: Lecture Notes in Computer Science (including
  subseries Lecture Notes in Artificial Intelligence and Lecture Notes in
  Bioinformatics), Vol. 6611 LNCS, Springer Verlag, 2011, pp. 338--349.
\newblock \href {http://dx.doi.org/10.1007/978-3-642-20161-5{\_}34}
  {\path{doi:10.1007/978-3-642-20161-5{\_}34}}.
\newline\urlprefix\url{http://query.nytimes.com/search/}

\bibitem{Hong2010EmpiricalTwitter}
L.~Hong, B.~D. Davison, \href{http://www.bit.ly}{{Empirical study of topic
  modeling in Twitter}}, in: SOMA 2010 - Proceedings of the 1st Workshop on
  Social Media Analytics, 2010, pp. 80--88.
\newblock \href {http://dx.doi.org/10.1145/1964858.1964870}
  {\path{doi:10.1145/1964858.1964870}}.
\newline\urlprefix\url{http://www.bit.ly}

\bibitem{Dixit2018}
S.~Dixit, S.~Kumar, P.~Kumar, \href{www.IJARIIT.com}{{Classification of tweets
  into various categories using classification methods}}, International Journal
  of Advance Research, Ideas and Innovations in Technology ISSN: 4~(3) (2018)
  937--939.
\newline\urlprefix\url{www.IJARIIT.com}

\bibitem{Alvarez-Melis2016TopicConversations}
D.~Alvarez-Melis, M.~Saveski, \href{www.aaai.org}{{Topic modeling in Twitter:
  Aggregating tweets by conversations}}, Tech. rep. (2016).
\newline\urlprefix\url{www.aaai.org}

\bibitem{deSouza2020AMedia}
J.~V. de~Souza, J.~Gomes, F.~M.~d. Souza~Filho, A.~M.~d. Oliveira~Julio, J.~F.
  de~Souza, \href{https://doi.org/10.1007/s13278-020-00659-2}{{A systematic
  mapping on automatic classification of fake news in social media}} (12 2020).
\newblock \href {http://dx.doi.org/10.1007/s13278-020-00659-2}
  {\path{doi:10.1007/s13278-020-00659-2}}.
\newline\urlprefix\url{https://doi.org/10.1007/s13278-020-00659-2}

\bibitem{Chatterjee2016TwitterCredibility}
R.~Chatterjee, S.~Agarwal, {Twitter truths: Authenticating analysis of
  information credibility}, in: 3rd International Conference on Computing for
  Sustainable Global Development (INDIACom), IEEE Conference Publication, New
  Delhi, 2016, pp. 2352--2357.

\bibitem{Vohra2018DetectionMedia}
M.~Vohra, M.~Kakkar, {Detection of Rumor in Social Media}, in: 2018 8th
  International Conference on Cloud Computing, Data Science {\&} Engineering
  (Confluence), 2018, pp. 485--490.
\newblock \href {http://dx.doi.org/10.1109/CONFLUENCE.2018.8442442}
  {\path{doi:10.1109/CONFLUENCE.2018.8442442}}.

\bibitem{Liu2019}
X.~Liu, {A big data approach to examining social bots on Twitter}, Journal of
  Services Marketing 33~(4) (2019) 369--379.
\newblock \href {http://dx.doi.org/10.1108/JSM-02-2018-0049}
  {\path{doi:10.1108/JSM-02-2018-0049}}.

\bibitem{YanZhang2016ADifferences}
{Yan Zhang}, W.~Chen, C.~K. Yeo, C.~T. Lau, B.~S. Lee,
  \href{http://ieeexplore.ieee.org/document/7936102/}{{A distance-based outlier
  detection method for rumor detection exploiting user behaviorial
  differences}}, in: 2016 International Conference on Data and Software
  Engineering (ICoDSE), IEEE, 2016, pp. 1--6.
\newblock \href {http://dx.doi.org/10.1109/ICODSE.2016.7936102}
  {\path{doi:10.1109/ICODSE.2016.7936102}}.
\newline\urlprefix\url{http://ieeexplore.ieee.org/document/7936102/}

\bibitem{Wu2018TowardMisinformation}
L.~Wu, J.~Li, F.~Morstatter, H.~Liu,
  \href{http://www.siam.org/journals/ojsa.php}{{Toward relational learning with
  misinformation}}, Tech. rep. (2018).
\newblock \href {http://dx.doi.org/10.1137/1.9781611975321.80}
  {\path{doi:10.1137/1.9781611975321.80}}.
\newline\urlprefix\url{http://www.siam.org/journals/ojsa.php}

\bibitem{Tambuscio2019Fact-checkingSociety}
M.~Tambuscio, G.~Ruffo,
  \href{https://doi.org/10.1007/s41109-019-0233-1}{{Fact-checking strategies to
  limit urban legends spreading in a segregated society}}, Applied Network
  Science 4~(1).
\newblock \href {http://dx.doi.org/10.1007/s41109-019-0233-1}
  {\path{doi:10.1007/s41109-019-0233-1}}.
\newline\urlprefix\url{https://doi.org/10.1007/s41109-019-0233-1}

\bibitem{Skolkay2019ASolutions}
A.~{\v{S}}kolkay, J.~Filin, \href{https://mediastudies.eu}{{A Comparison of
  Fake News Detecting and Fact-Checking AI Based Solutions}}, Studia
  Medioznawcze 20~(4) (2019) 365--383.
\newblock \href {http://dx.doi.org/10.33077/uw.24511617.ms.2019.4.187}
  {\path{doi:10.33077/uw.24511617.ms.2019.4.187}}.
\newline\urlprefix\url{https://mediastudies.eu}

\bibitem{Meyer2004ImpfgegnerImpfskeptiker}
C.~Meyer, S.~Reiter,
  \href{https://link.springer.com/article/10.1007/s00103-004-0953-x
  https://pubmed.ncbi.nlm.nih.gov/15583889/
  http://link.springer.com/10.1007/s00103-004-0953-x}{{Impfgegner und
  Impfskeptiker}}, Bundesgesundheitsblatt - Gesundheitsforschung -
  Gesundheitsschutz 47~(12) (2004) 1182--1188.
\newblock \href {http://dx.doi.org/10.1007/s00103-004-0953-x}
  {\path{doi:10.1007/s00103-004-0953-x}}.
\newline\urlprefix\url{https://link.springer.com/article/10.1007/s00103-004-0953-x
  https://pubmed.ncbi.nlm.nih.gov/15583889/
  http://link.springer.com/10.1007/s00103-004-0953-x}

\bibitem{Nguyen2020}
A.~Nguyen, D.~Catalan-Matamoros,
  \href{https://www.cogitatiopress.com/mediaandcommunication/article/view/3352}{{Digital
  mis/disinformation and public engagment with health and science
  controversies: Fresh perspectives from Covid-19}}, Media and Communication
  8~(2) (2020) 323--328.
\newblock \href {http://dx.doi.org/10.17645/mac.v8i2.3352}
  {\path{doi:10.17645/mac.v8i2.3352}}.
\newline\urlprefix\url{https://www.cogitatiopress.com/mediaandcommunication/article/view/3352}

\bibitem{Vraga2020}
E.~K. Vraga, M.~Tully, L.~Bode,
  \href{https://www.cogitatiopress.com/mediaandcommunication/article/view/3200
  https://www.standardmedia.co.ke/}{{Empowering users to respond to
  misinformation about covid-19}}, Media and Communication 8~(2) (2020)
  475--479.
\newblock \href {http://dx.doi.org/10.17645/mac.v8i2.3200}
  {\path{doi:10.17645/mac.v8i2.3200}}.
\newline\urlprefix\url{https://www.cogitatiopress.com/mediaandcommunication/article/view/3200
  https://www.standardmedia.co.ke/}

\bibitem{Schipani2020}
A.~Schipani, H.~Foy, J.~Webber, M.~Seddon,
  \href{https://www.ft.com/content/974dc9d2-77c1-4381-adcd-2f755333a36b}{{The
  ‘Ostrich Alliance’: the leaders denying the coronavirus threat |
  Financial Times}} (4 2020).
\newline\urlprefix\url{https://www.ft.com/content/974dc9d2-77c1-4381-adcd-2f755333a36b}

\bibitem{Friedman2020}
U.~Friedman,
  \href{https://www.theatlantic.com/politics/archive/2020/03/bolsonaro-coronavirus-denial-brazil-trump/608926/}{{Bolsonaro
  Leads the Coronavirus-Denial Movement - The Atlantic}} (3 2020).
\newline\urlprefix\url{https://www.theatlantic.com/politics/archive/2020/03/bolsonaro-coronavirus-denial-brazil-trump/608926/}

\bibitem{Larson2020}
H.~J. Larson, {Blocking information on COVID-19 can fuel the spread of
  misinformation}, Nature 580~(7803) (2020) 306--306.
\newblock \href {http://dx.doi.org/10.1038/d41586-020-00920-w}
  {\path{doi:10.1038/d41586-020-00920-w}}.

\bibitem{ARTICLE192020}
{ARTICLE 19},
  \href{https://www.article19.org/resources/brazil-civil-society-charges-government-at-iachr-for-violating-access-to-information-and-transparency-during-coronavirus-crisis/}{{Brazil:
  Civil society charges government with violating access to information and
  transparency during coronavirus crisis}} (7 2020).
\newline\urlprefix\url{https://www.article19.org/resources/brazil-civil-society-charges-government-at-iachr-for-violating-access-to-information-and-transparency-during-coronavirus-crisis/}

\bibitem{FolhadeS.Paulo2020}
{Folha de S. Paulo},
  \href{https://www1.folha.uol.com.br/internacional/en/scienceandhealth/2020/06/news-organizations-team-up-to-provide-transparency-to-covid-19-data.shtml}{{News
  organizations team up to provide transparency to covid-19 data - 08/06/2020 -
  Science and Health - Folha}} (6 2020).
\newline\urlprefix\url{https://www1.folha.uol.com.br/internacional/en/scienceandhealth/2020/06/news-organizations-team-up-to-provide-transparency-to-covid-19-data.shtml}

\bibitem{Batista2020}
J.~Batista,
  \href{https://ijnet.org/en/story/journalism-collective-fact-checks-religious-news-brazil}{{Journalism
  collective fact-checks religious news in Brazil}} (7 2020).
\newline\urlprefix\url{https://ijnet.org/en/story/journalism-collective-fact-checks-religious-news-brazil}

\bibitem{Dourado2020AgenciaINTERNET}
J.~L. Dourado, M.~T. Alencar,
  \href{https://orcid.org/0000-0003-1478-0012}{{Ag{\^{e}}ncia Lupa:
  fact-checking como modelo de neg{\'{o}}cio na Internet AG{\^{E}}NCIA LUPA:
  FACT-CHECKING AS A BUSINESS MODEL ON THE INTERNET}}, Comunica{\c{c}}{\~{a}}o
  {\&} Inova{\c{c}}{\~{a}}o 21~(46).
\newblock \href {http://dx.doi.org/10.13037/ci.vol21n46.6388}
  {\path{doi:10.13037/ci.vol21n46.6388}}.
\newline\urlprefix\url{https://orcid.org/0000-0003-1478-0012}

\bibitem{Moreno2019Factck.br:News}
J.~Moreno, G.~Bressan,
  \href{http://dl.acm.org/citation.cfm?doid=3323503.3361698}{{Factck.br: A new
  dataset to study fake news}}, in: Proceedings of the 25th Brazillian
  Symposium on Multimedia and the Web, WebMedia 2019, Association for Computing
  Machinery, Inc, New York, New York, USA, 2019, pp. 525--527.
\newblock \href {http://dx.doi.org/10.1145/3323503.3361698}
  {\path{doi:10.1145/3323503.3361698}}.
\newline\urlprefix\url{http://dl.acm.org/citation.cfm?doid=3323503.3361698}

\bibitem{IFCNInternationalPoynter}
{IFCN}, \href{https://www.poynter.org/ifcn/}{{International Fact-Checking
  Network – Poynter}}.
\newline\urlprefix\url{https://www.poynter.org/ifcn/}

\bibitem{Stencel2016GlobalLab}
M.~Stencel,
  \href{https://reporterslab.org/global-fact-checking-up-50-percent/}{{Global
  fact-checking up 50{\%} in past year - Duke Reporters' Lab}} (2 2016).
\newline\urlprefix\url{https://reporterslab.org/global-fact-checking-up-50-percent/}

\bibitem{Blei2003LatentJordan}
D.~M. Blei, A.~Y. Ng, J.~B. Edu, {Latent Dirichlet Allocation Michael I.
  Jordan}, Tech. rep. (2003).

\bibitem{Scott1990IndexingAnalysis}
D.~Scott, D.~S. T, F.~G. W, L.~T. K, H.~Richard,
  \href{papers2://publication/uuid/BA23C102-A2EA-4493-AEA4-2B85C3BD3FA1}{{Indexing
  by latent semantic analysis}}, Journal of the American Society for
  Information Science 41~(6) (1990) 391--407.
\newline\urlprefix\url{papers2://publication/uuid/BA23C102-A2EA-4493-AEA4-2B85C3BD3FA1}

\bibitem{Martin2015MoreApproach}
F.~Martin, M.~Johnson, \href{http://www.nltk.org/howto/wordnet.html}{{More
  Efficient Topic Modelling Through a Noun Only Approach}}, Tech. rep. (2015).
\newline\urlprefix\url{http://www.nltk.org/howto/wordnet.html}

\bibitem{Lu2012TrendTwitterb}
R.~Lu, Q.~Yang, {Trend Analysis of News Topics on Twitter}, International
  Journal of Machine Learning and Computing 2~(3) (2012) 327--332.
\newblock \href {http://dx.doi.org/10.7763/ijmlc.2012.v2.139}
  {\path{doi:10.7763/ijmlc.2012.v2.139}}.

\bibitem{Coviello2014ClusteringHEM}
E.~Coviello, A.~B. Chan, G.~R. Lanckriet, {Clustering hidden Markov models with
  variational HEM}, Tech. rep. (2014).

\bibitem{Rodrigues2014Lemport:Portuguese}
R.~Rodrigues, H.~G. Oliveira, P.~Gomes,
  \href{http://lxcenter.di.fc.ul.pt/services/en/LXServicesSuite.html.}{{Lemport:
  A high-accuracy cross-platform lemmatizer for portuguese}}, OpenAccess Series
  in Informatics 38 (2014) 267--274.
\newblock \href {http://dx.doi.org/10.4230/OASIcs.SLATE.2014.267}
  {\path{doi:10.4230/OASIcs.SLATE.2014.267}}.
\newline\urlprefix\url{http://lxcenter.di.fc.ul.pt/services/en/LXServicesSuite.html.}

\bibitem{Vermeer2020TowardApproach}
S.~Vermeer, D.~Trilling,
  \href{https://www.tandfonline.com/action/journalInformation?journalCode=rjos20}{{Toward
  a Better Understanding of News User Journeys: A Markov Chain Approach}},
  Journalism Studies 21~(7) (2020) 879--894.
\newblock \href {http://dx.doi.org/10.1080/1461670x.2020.1722958}
  {\path{doi:10.1080/1461670x.2020.1722958}}.
\newline\urlprefix\url{https://www.tandfonline.com/action/journalInformation?journalCode=rjos20}

\bibitem{Tierney1991ExploringChains}
L.~Tierney, {Exploring Posterior Distributions Using Markov Chains} (1991)
  563--570.

\bibitem{Geyer1992PracticalCarlo}
C.~J. Geyer, {Practical markov chain monte carlo}, Statistical Science 7~(4)
  (1992) 473--483.
\newblock \href {http://dx.doi.org/10.1214/ss/1177011137}
  {\path{doi:10.1214/ss/1177011137}}.

\bibitem{Yang2016ANetworks}
Z.~Yang, R.~Algesheimer, C.~J. Tessone, \href{www.nature.com/scientificreports
  http://www.nature.com/articles/srep30750}{{A comparative analysis of
  community detection algorithms on artificial networks}}, Scientific Reports
  6~(1) (2016) 30750.
\newblock \href {http://dx.doi.org/10.1038/srep30750}
  {\path{doi:10.1038/srep30750}}.
\newline\urlprefix\url{www.nature.com/scientificreports
  http://www.nature.com/articles/srep30750}

\bibitem{Lovari2020SpreadingItaly}
A.~Lovari,
  \href{https://www.cogitatiopress.com/mediaandcommunication/article/view/3219}{{Spreading
  (Dis)trust: Covid-19 misinformation and government intervention in Italy}},
  Media and Communication 8~(2) (2020) 458--461.
\newblock \href {http://dx.doi.org/10.17645/mac.v8i2.3219}
  {\path{doi:10.17645/mac.v8i2.3219}}.
\newline\urlprefix\url{https://www.cogitatiopress.com/mediaandcommunication/article/view/3219}

\bibitem{Nguyen2020Covid-19Perspective}
H.~Nguyen, A.~Nguyen,
  \href{https://www.cogitatiopress.com/mediaandcommunication/article/view/3227}{{Covid-19
  Misinformation and the Social (Media) Amplification of Risk: A Vietnamese
  Perspective}}, Media and Communication 8~(2) (2020) 444--447.
\newblock \href {http://dx.doi.org/10.17645/mac.v8i2.3227}
  {\path{doi:10.17645/mac.v8i2.3227}}.
\newline\urlprefix\url{https://www.cogitatiopress.com/mediaandcommunication/article/view/3227}

\bibitem{Statista2020Brazil:COVID-19}
{Statista},
  \href{https://www.statista.com/statistics/1118867/bolsonaro-fake-statements-coronavirus/}{{Brazil:
  Bolsonaro's fake statements on COVID-19}} (6 2020).
\newline\urlprefix\url{https://www.statista.com/statistics/1118867/bolsonaro-fake-statements-coronavirus/}

\bibitem{Pennycook2018PriorNews}
G.~Pennycook, T.~D. Cannon, D.~G. Rand, {Prior exposure increases perceived
  accuracy of fake news}, Journal of Experimental Psychology: General 147~(12)
  (2018) 1865--1880.
\newblock \href {http://dx.doi.org/10.1037/xge0000465}
  {\path{doi:10.1037/xge0000465}}.

\bibitem{Phillips2020BrazilSite}
D.~Phillips,
  \href{https://www.theguardian.com/world/2020/jun/07/brazil-stops-releasing-covid-19-death-toll-and-wipes-data-from-official-site}{{Brazil
  stops releasing Covid-19 death toll and wipes data from official site}} (6
  2020).
\newline\urlprefix\url{https://www.theguardian.com/world/2020/jun/07/brazil-stops-releasing-covid-19-death-toll-and-wipes-data-from-official-site}

\bibitem{Chan2017Debunking:Misinformation}
M.~p.~S. Chan, C.~R. Jones, K.~Hall~Jamieson, D.~Albarrac{\'{i}}n,
  \href{http://journals.sagepub.com/doi/10.1177/0956797617714579}{{Debunking: A
  Meta-Analysis of the Psychological Efficacy of Messages Countering
  Misinformation}}, Psychological Science 28~(11) (2017) 1531--1546.
\newblock \href {http://dx.doi.org/10.1177/0956797617714579}
  {\path{doi:10.1177/0956797617714579}}.
\newline\urlprefix\url{http://journals.sagepub.com/doi/10.1177/0956797617714579}

\end{thebibliography}
\end{document}